\documentclass[sigconf,authorversion,nonacm]{acmart}
\AtBeginDocument{%
  \providecommand\BibTeX{{%
    \normalfont B\kern-0.5em{\scshape i\kern-0.25em b}\kern-0.8em\TeX}}}

\setcopyright{acmlicensed}
\copyrightyear{2024}
\acmYear{2024}
\acmDOI{10.1145/3589335.3651959}

\acmConference[WWW '24 Companion]{Companion Proceedings of the ACM Web Conference 2024}{May 13--17, 2024}{Singapore, Singapore}
\acmBooktitle{Companion Proceedings of the ACM Web Conference 2024 (WWW '24 Companion), May 13--17, 2024, Singapore, Singapore}

\acmISBN{979-8-4007-0172-6/24/05}

\usepackage{tabularx} 
\usepackage{makecell}
\usepackage{placeins}

\begin{document}

\title{Detecting Financial Bots on the Ethereum Blockchain}

\author{Thomas Niedermayer}
\email{thomas.niedermayer@ikna.io}
\orcid{0009-0006-8464-8542}
\affiliation{%
  \institution{Iknaio Cryptoasset Analytics}
  \city{Vienna}
  \country{Austria}
}
\author{Pietro Saggese}
\email{pietro.saggese@imtlucca.it}
\orcid{0000-0002-7387-0846}
\affiliation{%
  \institution{IMT School for Advanced Studies}
  \city{Lucca}
  \country{Italy}
}

\author{Bernhard Haslhofer}
\email{haslhofer@csh.ac.at}
\orcid{0000-0002-0415-4491}
\affiliation{%
  \institution{Complexity Science Hub}
  \city{Vienna}
  \country{Austria}
}

\renewcommand{\shortauthors}{Niedermayer et al.}


\begin{abstract}

The integration of bots in Distributed Ledger Technologies (DLTs) fosters efficiency and automation.
However, their use is also associated with predatory trading and market manipulation, and can pose threats to system integrity. It is therefore essential to understand the extent of bot deployment in DLTs; despite this, current detection systems are predominantly rule-based and lack flexibility. 
In this study, we present a novel approach that utilizes machine learning for the detection of financial bots on the Ethereum platform. First, we systematize existing scientific literature and collect anecdotal evidence to establish a taxonomy for financial bots, comprising 7 categories and 24 subcategories. Next, we create a ground-truth dataset consisting of 133 human and 137 bot addresses. Third, we employ both unsupervised and supervised machine learning algorithms to detect bots deployed on Ethereum. The highest-performing clustering algorithm is a Gaussian
Mixture Model with an average cluster purity of 82.6\%, while the highest-performing model for binary classification is a Random Forest
with an accuracy of 83\%. Our machine learning-based detection mechanism contributes to understanding the Ethereum ecosystem dynamics by providing additional insights into the current bot landscape.

\end{abstract}




\received{20 February 2007}
\received[revised]{12 March 2009}
\received[accepted]{5 June 2009}

\maketitle

\section{Introduction}

The algorithmic automation of financial activity is among the most fundamental innovations harnessed by Decentralized Finance (DeFi), and it is integral to achieving efficiency within this framework \cite{auer_technology_2023}.
In Distributed Ledger Technologies (DLTs) like Ethereum, 
bots facilitate the automated execution of software programs, known as smart contracts, that encapsulate the logic of deterministic functions such as conventional financial operations. 
Bots typically manage externally owned accounts (EOAs), often referred to as wallets, which are identified by a hexadecimal address. As there is significant incentive to automate EOAs in the form of financial gain, time saving, or increased fault tolerance, bots have emerged as impactful agents on DLTs, and their relevance has become more prominent over time. 
For instance, bots have been increasingly exploited to extract more than \$1 billion profits from the Ethereum ecosystem to date \cite{qin_quantifying_2021, auer_miners_2022, flashbots_flashbots_2023}.

The deployment of bots can be beneficial to DLT-based ecosystems. Bots can be used to provide critical infrastructure, e.g., enabling interfaces between centralized exchanges and the blockchain by managing assets automatically \cite{saggese_assessing_2023}. Furthermore, layer 2 scaling solutions like roll-ups automatically deploy transaction information on Ethereum for persistence \cite{thibault_blockchain_2022}. Bots also contribute to market stability through arbitrage \cite{wang_cyclic_2022}.

However, bots are also exploited for predatory trading and market manipulation~\cite{qin_quantifying_2021}, posing financial threats to unwary users \cite{wang_impact_2022} and potential systemic threats to the network's integrity \cite{barczentewicz_mev_2023}. 
In cases of profit-optimizing bots, side-effects on the ecosystem can be the exploitation of users interacting with smart contracts \cite{torres_frontrunner_2021}, exploitation of smart contracts \cite{zhang_chi-researchsymbolic-searcher_2023}, inflated gas prices through increased traffic, and the destabilization of the network \cite{daian_flash_2020}.

Whilst understanding the scale of bot adoption in DLTs is essential for assessing risks to human users, ensuring their protection, and informing policy, it is hard to estimate the extent of this phenomenon, as to date there are no dedicated bot detection systems. Existing software such as MEV-inspect\footnote{\url{https://github.com/flashbots/mev-inspect-py}} and Eigenphi\footnote{\url{https://eigenphi.io/}}, that focus on profit-optimizing transactions often carried out by bots, may be repurposed for bot detection. However, they are rule-based and are not built for capturing evolving bot behaviors. %
It is therefore important to devise bot detection methods that are more flexible and operate in a data-driven manner. %

\begin{figure*}[t]
  \includegraphics[width=0.9\textwidth]{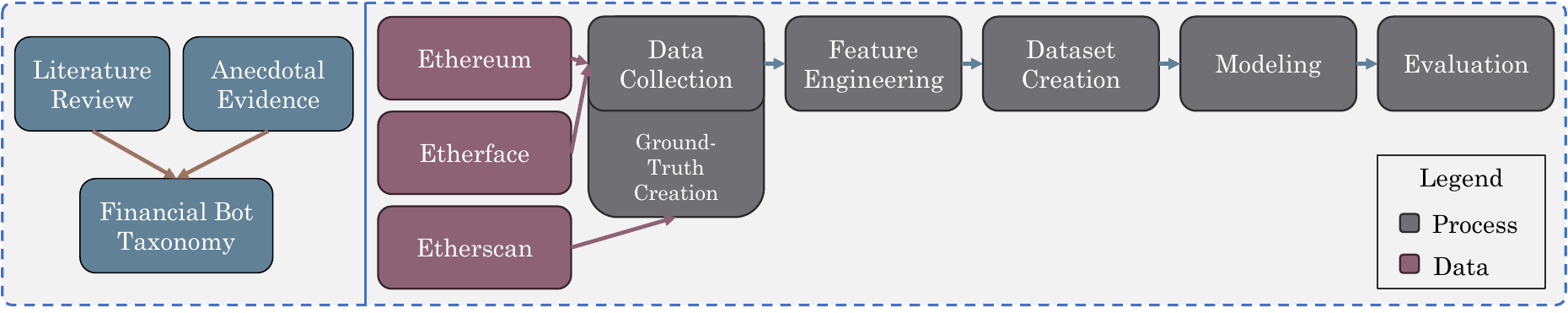}
  \caption{Overview of the methodology.}
  \label{fig:overview}
  \Description{Pipeline of the analysis. The data from Erigon, Etherface and Etherscan goes into the Data Collection Process followed by the processes feature engineering, dataset creation, modeling, and evaluation in that order.}
\end{figure*}

This study provides a machine learning-based method to detect financial bots deployed on Ethereum and aims to identify key bot predictors, in an effort to remove the prevalent reliance on mechanism-specific and rule-based systems.
Figure \ref{fig:overview} provides an overview of the pipeline of our work. First, we complement existing literature with anecdotal evidence to propose a taxonomy for financial bots on Ethereum. Second, we process publicly available Ethereum blockchain data and additional information from the websites Etherface.io and Etherscan.io, and we create a ground-truth dataset by annotating 270 EOAs based on the assessment of three independent annotators. We then extract features to accurately represent EOA behavior for testing bot detection using machine learning algorithms. Finally, we evaluate the effectiveness of these models. We utilize both unsupervised and supervised techniques to respectively group and classify EOAs as either ``Bot'' or ``Human''. Additionally, we focus on a specific 
subset of asset-accumulating bots, and measure algorithm performance in a multiclass classification context.
The research questions and related contributions of this work can be summarized as follows:

\begin{enumerate}

    \item[RQ1] \textbf{What types of bots are active on distributed ledger technologies such as Ethereum?} We propose a taxonomy for financial bots on Ethereum comprising 7 bot categories and 24 subcategories. Furthermore, we create a ground-truth dataset of 133 human and 137 bot addresses.

    \item[RQ2] \textbf{To what extent can we detect bots automatically?} We deploy both supervised and unsupervised machine learning models and measure their performance. The highest-performing clustering algorithm is a Gaussian Mixture Model with an average cluster purity of 82.6\%, evaluated on a balanced two-class dataset. The highest-performing binary classification model is a Random Forest with an accuracy of 83\% on the same dataset.

    \item[RQ3] \textbf{Which features are most informative to a model with high performance?} Using explainable AI techniques, we find that features based on the time, frequency, gas price, and gas limit of outgoing transactions are the most influential.

\end{enumerate}

We believe that our work helps better understand the Ethereum ecosystem dynamics by shedding more light on the existing bot landscape and by proposing a novel ML-based detection mechanism.
As these bots can significantly influence market dynamics, liquidity, and the overall network safety, it is important to monitor their prevalence and impact.

All data and source code used in this study are publicly available for reproducibility purposes\footnote{\url{https://github.com/Tommel71/Ethereum-Bot-Detection}}.

\section{Background}
This section provides a background of the technologies used by financial bots, defines what a bot is in a blockchain context, and summarizes the existing literature on bot detection.

\subsection{Decentralized Finance (DeFi)}
Ethereum's smart contract support allows the execution of programs in a decentralized manner. This gives rise to complex functionalities involving cryptoassets, i.e., digital assets that represent some economic resource or value to someone and are based on cryptographic primitives and a DLT \cite{auer_technology_2023}.
Decentralized Finance (DeFi) enables trustless borrowing, lending, or investing on blockchain systems like Ethereum \cite{werner_sok_2023}. The following are core components of DeFi relevant to this work.

Tokens\footnote{\url{https://ethereum.org/en/developers/docs/standards/tokens}} are cryptoassets created through smart contracts that adhere to specific standards to implement core functionalities such as the transferability between accounts \cite{vogelsteller_erc-20_nodate}.
Tokens following the ERC-20 standard are fungible, i.e. each unit is identical to any other deployed by the same contract. Smart contracts implemented according to the ERC-721 standard are Non-Fungible Tokens (NFTs). This means they are not interchangeable, which makes each NFT unique, allowing them to be used as proof of ownership of a digital asset. Tokens have been used to represent Dollar equivalents on a blockchain (stablecoins) and NFTs have been used to represent artworks, tickets, or characters in video games \cite{wang_non-fungible_2021}. %

Decentralized exchanges (DEXes) on Ethereum are smart contracts that create a market to exchange cryptoassets against each other. Their off-chain counterparts, i.e. centralized exchanges (CEX), are commonly used as an interface between fiat currencies and cryptoassets and often enable trading through an order book. DEXes on the other hand often employ Automated Market Makers (AMMs), that allow for trading without an order book \cite{mohan_automated_2022}. For these AMMs to work, prevalent mechanisms require liquidity to enable trading and users are financially incentivized to provide it \cite{lehar_decentralized_2021}.

A relevant concept in DeFi is Maximal Extractable Value (MEV), initially termed miner-extractable value in the seminal work of Daian et al.~\cite{daian_flash_2020}. It refers to profits that can be extracted from block production in excess of the standard block reward and transaction fees. This is achieved by including, excluding, or re-ordering transactions within a block \cite{smith_maximal_2023}. The most prominent examples of MEV are arbitrage, sandwich attacks, and liquidations. Arbitrage exploits price differences on DEXes to generate profit. Sandwich attacks send a transaction to drive up the price, let the victim buy at the high price, to then send another transaction to sell what was bought earlier at a profit. Liquidations involve the sale of collateralized assets to repay a debt at a favorable rate for the buyer \cite{qin_empirical_2021}.

\subsection{Bots}
The potential for automation in Ethereum's ecosystem and the amount of assets involved in DeFi attract a diverse set of actors categorized as bots. In the context of account-based blockchain systems such as Ethereum, Zwang et al. \cite{zwang_detecting_2018} defined bots as software robots. In this work, we describe them more precisely by defining the two mutually exclusive classes \textit{human} and \textit{bot} and establish the following Definition \ref{def:bot}:

\begin{definition}[Bot]
\label{def:bot}
A bot is an EOA that has sent at least one transaction that was compiled and sent by a software program without human oversight, i.e. a bot transaction.
\end{definition}

\subsection{Bot Detection and Related Literature}
\label{sec:bg_lit}
Software programs such as MEV-inspect and Eigenphi can discern transactions related to MEV. As these transactions are highly time-critical, bots are employed to execute them; we can therefore interpret these programs as bot detection systems, by recognizing the sender of detected transactions as bots. Currently, these systems primarily focus on transaction patterns in a rule-based manner and are focused on a specific set of transactions modeled in their code.
In addition to such programs, research related to the identification of bots on Ethereum has been conducted. Studies leverage heuristic methods, which allow to analyze bots without expensive annotation efforts. Graph-based heuristics have shed light on bot types and the prevalence of private transaction use in MEV exploitation \cite{piet_extracting_2022}. Time-based heuristics have also been instrumental in identifying automated wallet behaviors, revealing patterns in transaction timings \cite{zwang_detecting_2018}. Further, a categorization of Ethereum wallets using transaction-based and graph-theoretic metrics has been carried out \cite{bonifazi_defining_2022}. An analysis of bots involved in sniping, a subset of MEV that aims to acquire cryptoassets as fast as possible to profit from other users driving up the price, highlights their operational tactics and prevalence through heuristic examination of transaction timings \cite{cernera_ready_2023}. Overall, these heuristic-based studies provide a solid basis for bot detection and motivate features that could be implemented in a more flexible ML-based system.

MEV bots have received particular attention because of MEV's impact on Ethereum security and economy, as highlighted by recent research \cite{qin_quantifying_2021, flashbots_flashbots_2023}. Because time is critical to exploit MEV opportunities, bots are employed to automatically find them and assemble and send transactions that act upon them. Key studies have shown the practical threats posed by MEV, focusing on real-time pending transactions data analysis \cite{daian_flash_2020}. There is a systematization of knowledge around front-running on blockchains, drawing parallels with traditional finance and classifying front-running based on adversarial intent \cite{eskandari_sok_2020}.

Whilst we are not aware of any work that uses ML for bot detection specifically, several works used ML approaches for the broader task of behavior-based wallet classification. Unsupervised learning, specifically using expectation-maximization algorithms coupled with Random Forest models, has been applied for clustering Ethereum wallets and detecting anomalies \cite{baek_model_2019}. Key statistical measures have been calculated over data grouped by Ethereum addresses and effectively used for machine learning-based classification of accounts \cite{baek_model_2019, galletta_sharpening_2023}. 
Supervised and unsupervised techniques have been combined to classify attack instances, utilizing deep neural networks for attack detection \cite{rabieinejad_deep_2021}. Ponzi scheme detection in Ethereum, using a combination of manual labeling and feature extraction, exemplifies the application of machine learning in identifying fraudulent activities \cite{chen_exploiting_2019, galletta_sharpening_2023}. Additionally, anomaly detection in Bitcoin transactions, employing methods like one-class support vector machines and k-means clustering have been surveyed \cite{sayadi_anomaly_2019}.

In conclusion, we found software applicable for bot detection, research employing rule-based heuristics to detect bots, and more generally, research in the use of ML to detect certain behaviors of accounts. To the best of our knowledge, however, there is no research on ML-based bot detection on Ethereum. This serves as a rationale to motivate our study.%

\section{Bot Categories}

\begin{figure}[t]
  \includegraphics[width=0.9\linewidth]{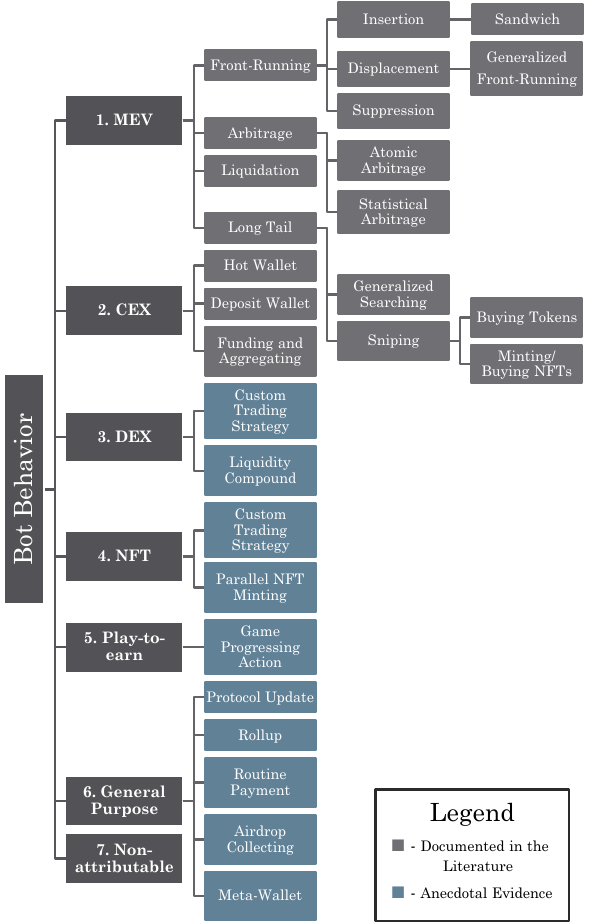}
  \caption{Taxonomy of Bots on Ethereum.}
  \label{fig:taxonomy}
  \Description{Taxonomy of Bots on Blockchains.}
\end{figure}

In an effort to provide a comprehensive representation of the DeFi bot landscape in Ethereum, we complement the existing scientific literature with anecdotal evidence of existing financial bots that we collected from GitHub repositories, sample transactions, or sample accounts (see Table \ref{tab:bot_evidence} in Appendix for further details). 
We report our taxonomy in Figure \ref{fig:taxonomy}, divided into 7 categories and 24 subcategories, based on the collected evidence. Below we outline the characteristics of the seven categories identified. 

\textit{MEV bots} exploit a blockchain state for profit by rapidly detecting and executing profitable opportunities, raising concerns about fairness and protocol integrity.
\textit{Centralized Exchange (CEX) bots} handle interactions between centralized exchanges and blockchains. They include hot wallets for daily operations, deposit wallets that manage user funds, and funding wallets that provide gas-fee-covering liquidity.
\textit{Decentralized Exchange (DEX) bots} engage in custom trading strategies and liquidity compounding. They use automated mechanisms to trade tokens or provide liquidity to pools, often for staking rewards.
\textit{Non-Fungible Token (NFT) bots} engage in trading strategies and minting practices. This includes automated trading based on pricing models and parallel minting to circumvent bulk limits, imposed by smart contracts to prevent exploitation.
\textit{Play-to-Earn (P2E) bots} automate game-progressing actions in blockchain-based games for rewards. Common actions include questing, combat, crafting, and market interactions. Automated processes enable constant game progression and asset accumulation.
\textit{General Purpose bots} execute routine tasks like protocol updates, rollups, payments, and airdrop collecting. They automate repetitive or resource-intensive tasks. Finally, \textit{Non-Attributable bots} exhibit automated behavior without a clear purpose and are therefore categorized as non-attributable.
\section{Data and Methods}
In the previous sections, we have defined bots and given an overview of the bot landscape. We now describe how we process publicly available blockchain data and smart contract information to create our datasets.

\subsection{Data Sources}
\label{data_sources}
We gathered data from the following sources:

\par \textbf{Ethereum blockchain:} We ingest transaction data from an Erigon archive node\footnote{\url{https://erigon.tech/}} using \textit{Graphsense-lib}\footnote{\url{https://github.com/graphsense/graphsense-lib}} \cite{haslhofer_graphsense_2021}. We investigate the block range of 15,500,000 (September 9, 2022) until 15,599,999 (September 24, 2022), i.e. 100,000 blocks. 
This limited range allows our method to run on consumer hardware.
Furthermore, the range is defined so that it contains no significant hype-phase which might bias the analysis, and all major features of DeFi as we know it are already established, which is relevant for many bot categories. While the method described here uses 100,000 blocks, the block range can be extended or restricted arbitrarily to include more information in the features or allow for faster processing, respectively. 
\par
\par \textbf{Etherface.io:} We use smart contract function and event descriptions from Etherface.io\footnote{\url{Etherface.io}} to decode smart contract functions and events and make data in swaps and token transfers accessible (see Table \ref{tab:decoded_functions} in the Appendix).
\par \textbf{Etherscan.io:} Using the API of Etherscan.io\footnote{\url{https://etherscan.io/}}, we provide additional transaction data about addresses annotated to allow annotators a more comprehensive analysis.

\subsection{Ground-Truth Creation}
\label{ground_truth_creation}
We call the last two blocks of our block range the test blocks and annotate them to yield a labeled dataset. 
To manually annotate all 270 EOAs sending transactions in the test blocks, we employ two independent annotators A and B. Based on information documented on the web, indexing services, and custom aggregate metrics and graphs, both annotators attach to all EOAs one of the labels ``Bot'' and ``Human''. In addition to the binary labels, annotator A adds fine-grained labels, indicating the specific use of a bot\footnote{We did not provide the annotators with a preset list of bot categories to prevent limiting them to our initial findings.  Additionally, our annotated block range is too small to cover all classes in the taxonomy. While these circumstances partly disconnect our bot taxonomy from the fine-grained labels, there is still notable overlap.}. In the case of a disagreement, a third annotator C breaks the tie based on a description noted down by annotators A and B. In the case of a tie-breaker where A picks Human (and therefore provides no fine-grained bot label) and C chooses B, then B provides the fine-grained label.
The result is a dataset of 137 bot and 133 human addresses, where 31 addresses required a tie-breaker. The Cohen’s Kappa of the first two annotators is 0.77, which indicates substantial agreement. Finally, to enable the analysis of a classifier discerning multiple bot classes, we gather additional data about short-tail MEV bots. We run MEV-inspect on the entire block range to obtain 111 bots for each category \textit{Arbitrage}, \textit{Sandwich}, and \textit{Liquidation} bots. The number of bots is the same for all categories by design and is limited by the lowest number of bots found in the category \textit{Liquidation}.

\subsection{Feature Engineering}
We perform feature engineering by transforming the transaction history of an address into aggregate metrics. This is mostly done for outgoing transactions or token transfers and in some cases for incoming ones. We calculate a total of 83 features per wallet. These metrics are in some cases informed by knowledge of trading bots or human behavior while in other instances, they reflect the address' overall activity to provide a subsequent model with highly informative features. Figure \ref{fig:features} in the Appendix shows an overview of the features, which are divided into four main categories: address-based features (based on the 40-character-long hexadecimal representation of an address and requires no transaction history), transaction-based features (based on transaction data), function-call-based features (based on input data sent in transactions to smart contracts), and event-based features (based on data smart contracts produce). The general strategy is to first extract specific data from the transaction history of an account to subsequently apply aggregate functions that produce one or more features. We first present such functions that are used multiple times, and then describe the features, which may require specific functions directly defined with the feature.

\par
\textit{General Functions.}
The function \textit{BenfordsLaw}, applied on a given set of values, leverages Benford’s law, which predicts the distribution of the first significant digits in natural datasets. 
The function applies Pearson's Chi-squared test to assess how well first-significant-digit distributions align with Benford’s law, generating a p-value as a feature \cite{benford_law_1938, cong_crypto_2021}.
To capture the tendency of human traders to use round numbers as cognitive reference points we define 
\textit{TradeValueClustering} as the ratio of round and non-round numbers in the grouped data \cite{cong_crypto_2021}. We call a number round if its digits after the 7th significant digit are zero. 
\textit{GapBasedSleepiness} calculates time deltas for a list of timestamps \( \mathrm{\textit{ts}} = [t_1, t_2, \dots, t_k] \) as \( \mathrm{\textit{TimeDelta(ts)}} = [t_2-t_1, t_3-t_2, \dots, t_k - t_{k-1}] \). Timestamps are divided into two-day intervals to minimize time-zone effects and account for human sleep patterns. In each interval, the maximum time delta is determined and then averaged across intervals to produce a single value.
\textit{Categorical} takes a group of categorical values and returns the entropy of the 
distribution of class occurrences, for each class the percentage of its occurrence, and the mode.
\textit{Numerical} takes a group of numerical values and returns mean, mode, standard deviation, minimum, maximum, and the 95\% quantile.

\subsubsection{Address-based features} The feature \textit{NLeadingZeros} counts leading zeros in Ethereum's hexadecimal addresses. Addresses with more leading zeros, often mined for efficiency, indicate deliberate creation to reduce gas fees and increase smart contract efficiency \cite{0age_efficient_2019}. \textit{DigitEntropy} measures the entropy of digits in an Ethereum address, capturing randomness or patterns which could be expressed differently in bots and therefore contain predictive power.

\subsubsection{Transaction-based features}
\textit{TxPerBlock} calculates the number of transactions sent divided by the total number of blocks (100,000).
\textit{StatisticsTxPerActiveBlock} calculates the number of transactions sent for each block, removes zeros, and applies \textit{Numerical}. The resulting metrics inform about the intensity of activity and its distribution in blocks where transcations were sent.

For time-based features, \textit{GapBasedSleepiness} is applied to the timestamps of in- and outgoing transactions separately. \textit{TransactionFrequency} calculates for both in- and outgoing transactions separately the total number of transactions divided by the number of seconds between first and last transaction made. \textit{StatisticsTimeDifference} represents the features returned by the \textit{Numerical} function applied to the group of time differences between consecutive transactions sent. \textit{EntropyTime} is done for both in- and outgoing transactions separately and calculates the entropy (see Appendix \ref{sec:metrics}) of the timestamps after transforming them to represent the hours \(h \in [1, 2,\dots,2 3, 24]\) of the day they were sent in \cite{liu_feature_2018}. For value-based features, \textit{BenfordsLaw} and \textit{TradeValueClustering} are calculated for the ETH values transferred in outgoing transactions. \textit{Standard} fields are the categorical ``type'' and ``status'', and the numerical ``value'', ``gasLimit'', and ``gasPrice'', together with the index of a transaction\footnote{All these values are found in Erigon transaction and receipt data.} normalized by the number of transactions in the block named ``indexRelative''. For each data group of a standard field, \textit{Categorical} or \textit{Numerical} is applied according to the data type.

\subsubsection{Function-Call-based features}
To filter blockchain data for specific smart contract function calls or emitted events related to DEXs, we modeled them based on the Uniswap protocol because it is pioneering the field of DEXs with an open-source code base. This has led to many other protocols copying Uniswap's code and interfaces allowing us to capture their functionality by modeling functions and events of Uniswap only  (see Table \ref{tab:decoded_functions}). 

We filter the outgoing transaction history for swaps where we modeled the contract. Swaps are especially relevant for financial bots as they represent a basic functionality many of them use and can therefore help in creating better representations of account behavior in the form of features. We calculate \textit{BenfordsLaw} and \textit{TradeValueClustering} for the ``amountIn'' or ``amountOut'' field denominated in the swap because in the case of a human using the website of a DEX, these fields are provided by the user while other fields in the smart contract call are often automatically calculated by the website before the transaction is sent. Automatically calculated values are less suited for metrics based on cognitive rules like \textit{TradeValueClustering}. \textit{StatisticsPathLength} takes for all outgoing transactions the path parameters which determines if the input asset is swapped directly into the output asset or if there are other swaps between, takes their length (two for a direct swap), and applies the \textit{Numerical} function. Finally, \textit{SwapsPerBlock} calculates the number of swap transactions divided by the total number of blocks.

\subsubsection{Event-based features}
For swaps based on UniswapV2, we get the input amount by adding up the fields ``amount0In'' and ``amount1In'' and for UniswapV3, we get the input amount by taking the maximum of the fields ``amount0'' and ``amount1''. We group the data by the recipient of the swap and not the sender in the event data because the senders are usually smart contracts. We calculate \textit{BenfordsLaw} and \textit{TradeValueClustering} for the input amount denominated in the swaps. Additionally, in \textit{SwapsPerBlock}, the number of swap events divided by the total number of blocks is given. Similarly, \textit{BenfordsLaw} and \textit{TradeValueClustering} are calculated for the value of transfer events with the standard ERC-20 transfer event signature. Finally, \textit{TransferEventsPerBlock} gives the number of transfer events divided by the total number of blocks.

For features based on swaps, we carry out a dimensionality reduction step that averages the respective features over all the functions and events modeled to reduce 128 features to 14. StatisticsPathLength features are only averaged over the functions modeled.

\subsection{Dataset Creation}
Lastly, we use the 100,000 blocks described in Section \ref{data_sources} to create three datasets of different sizes with each row representing one address as a set of features.

\begin{enumerate}
    \item \textbf{Binary Bot Dataset}: We collect all 270 addresses sending in the test blocks to produce this dataset. As described in Section \ref{ground_truth_creation}, these addresses are associated with the binary labels ``Bot'' and ``Human''.
    \item \textbf{Clustering Dataset}: We collect all addresses sending transactions in the 100,000 blocks window and exclude the ones sending in the test blocks. This dataset is unlabeled.
    \item \textbf{Multiclass MEV Dataset}: We randomly sample 111 addresses from the test blocks that are not involved with MEV and combine them with 111 addresses of each short-tail MEV category from the automatically annotated addresses. This results in a labeled dataset with 4 classes, each of size 111 with the corresponding labels ``Arbitrage'', ``Sandwich'', ``Liquidation'', and ``non-MEV''.
    
\end{enumerate}

\section{Results}\label{sec:results}
In this section, we analyze the effectiveness of clustering and classification to detect bots on Ethereum. Furthermore, we investigate which features are most influential in the decision of our best performing classifier.

\subsection{Clustering}
We begin by using clustering techniques to detect bots on Ethereum and explore how well bots and humans can be grouped in an unsupervised environment. To assess the quality of a clustering, we measure entropy and purity (see Appendix \ref{sec:metrics}) with
respect to the two classes Bot and Human in all individual clusters. Purity is a measure of homogeneity and a higher value means the cluster is more dominated by data points from a single class. Entropy is a measure of disorder in a cluster and is one for equal representation of all classes and zero for a cluster containing only one class. Therefore, for this analysis, we first utilize the unlabeled clustering dataset for fitting and next employ the binary bot dataset to compute purity and entropy.

After surveying multiple clustering methods, k-means and Gaussian Mixture Model (GMM) were chosen because they are inductive and can therefore be used to cluster new instances in the binary bot dataset, and because they run acceptably fast.
For both algorithms we use the scikit-learn \cite{pedregosa_scikit-learn_2011} implementation, whereby the number of clusters has to be provided. We tried different approaches to find the optimal number of clusters, such as minimization and the elbow method \cite{marutho_determination_2018} on the Bayesian Information Criterion (BIC) \cite{schwarz_estimating_1978}, and the silhouette score \cite{rousseeuw_silhouettes_1987}. While for k-means, minimizing the BIC yields more than 100 clusters and using the elbow method on BIC or silhouette scores yields 5 to 10 clusters, minimizing BIC for GMM yields 10 to 30 clusters. This wide range of cluster sizes suggested by common methods stands in conflict with the evaluation metrics we use. Since we have annotated data, we can calculate purity and entropy regarding the distribution of Bot and Human addresses within a cluster. These are appealing metrics because they tell whether or not a certain cluster can be used for selecting a higher percentage of either Bots or Humans. The conflict arises since a higher number of clusters, all else being equal, leads to higher purity because more clusters allow for fewer addresses per cluster on average, which in turn leads to a higher chance that the distribution in a cluster is pure just by chance. This effect becomes evident when a cluster may fit around one single address which leads to a purity of 100\%. For this reason, we evaluate the algorithms at three fixed cluster sizes of 5, 15, and 30, to obtain results at different levels of model complexity. 
Additional results based on BIC optimization and the `elbow' method are reported in the Appendix, Table~\ref{tab:clustering_results_purity_elbow}.

For both GMM and k-means, we preprocess the features in the clustering dataset by Min-Max-scaling to the \([0, 1]\) range, ensuring equal importance across features. We also examine whether UMAP embedding \cite{mcinnes_umap_2020} to two dimensions enhances performance and if imputing missing values with the column mean or -1 yields better outcomes. Imputing missing values with -1 post-scaling might be advantageous, as it distinctly separates addresses with missing values from those without in the feature space.

Finally, to calculate purity and entropy, all combinations of preprocessing methods, clustering methods, and number of clusters are fitted on the clustering dataset and tested by assigning clusters to the addresses in the binary bot dataset. 
For an entire clustering, we define the two metrics as the weighted average of the metrics of the individual clusters, weighted by the ratio of the samples in the respective cluster. 

\begin{table}[!h]
\caption{Average entropy and purity of the combinations of preprocessing methods and clustering algorithms explored with different cluster sizes.}
\label{tab:clustering_results_purity_clustersizefixed}
\resizebox{\columnwidth}{!}{

\begin{tabular}{lllrrrrrr}
\toprule
       &      &          &  \thead{Purity \\5 Clusters} &  \thead{Purity \\15 Clusters} &  \thead{Purity \\30 Clusters} &  \thead{Entropy \\5 Clusters} &  \thead{Entropy \\15 Clusters} &  \thead{Entropy \\30 Clusters} \\
Algorithm & Imputation & \thead{Dimension. \\Reduction} &                                            &                                             &                                             &                                             &                                              &                                              \\
\midrule
GMM & -1 & UMAP &                                      \textbf{0.726} &                                       0.744 &                                       0.756 &                                       0.842 &                                        0.795 &                                        \textbf{0.749} \\
       &      & non-UMAP &                                      0.581 &                                       0.670 &                                       0.789 &                                       0.934 &                                        0.791 &                                        0.606 \\
       & mean & UMAP &                                      0.544 &                                       0.741 &                                       0.770 &                                       \textbf{0.972} &                                        0.774 &                                        0.724 \\
       &      & non-UMAP &                                      0.552 &                                       0.770 &                                       \textbf{0.826} &                                       0.954 &                                        0.725 &                                        0.563 \\
kmeans & -1 & UMAP &                                      0.704 &                                       0.737 &                                       0.770 &                                       0.868 &                                        0.796 &                                        0.738 \\
       &      & non-UMAP &                                      0.633 &                                       0.674 &                                       0.711 &                                       0.906 &                                        \textbf{0.801} &                                        0.742 \\
       & mean & UMAP &                                      0.652 &                                       0.748 &                                       0.778 &                                       0.923 &                                        0.777 &                                        0.700 \\
       &      & non-UMAP &                                      0.585 &                                       \textbf{0.774} &                                       0.800 &                                       0.970 &                                        0.713 &                                        0.649 \\
\bottomrule
\end{tabular}

}
\end{table}

Table \ref{tab:clustering_results_purity_clustersizefixed} shows the results for different algorithms, preprocessing methods, and cluster sizes. GMM with 30 clusters, mean-imputation and no dimensionality reduction scores best in entropy and purity; we name it the top clustering algorithm. In general, UMAP-embedding is beneficial for a small number of clusters and detrimental in runs with large numbers of clusters. For smaller models, UMAP embedding even resulted in the best-performing clusterings. This could be due to embedding losing information that could be leveraged by larger models while helping to separate clusters effectively only using a few clusters by simplifying the data.

\begin{table}[!h]
\caption{Individual-cluster-level metrics of the largest clusters of the top clustering algorithm.}
\label{tab:clustering_results_best_combination_purity_entropy_clustersizefixed}

\begin{tabular}{lrrrl}
\toprule
{} &  Purity &  Entropy &  Size &        Majority \\
Cluster &         &          &       &                 \\
\midrule
2       &   0.667 &    0.918 &    18 &           Human \\
5       &   0.927 &    0.376 &    55 &             CEX \\
6       &   0.842 &    0.629 &    19 &           Human \\
13      &   0.929 &    0.371 &    28 &  Deposit Wallet \\
14      &   0.579 &    0.982 &    19 &           Human \\
19      &   0.872 &    0.552 &    39 &           Human \\
\bottomrule
\end{tabular}

\end{table}

Table \ref{tab:clustering_results_best_combination_purity_entropy_clustersizefixed} reports the 6 clusters produced by the top clustering algorithm that contain at least 15 addresses. The majority class within a cluster is based on the fine-grained label of the binary bot dataset. Clusters 6 and 19 have the majority class ``Human'' and show high purity. Clusters 5 and 13 also show exceptionally high purity with the majority class CEX and Deposit Wallet respectively. Clusters 2 and 14 have low quality, identifiable by an entropy close to 100\%.
Conclusively, there are easily discerned fine-grained bot classes such as Deposit Wallet and CEX, and with 30 clusters, a purity of 82.6\% can be attained.

\subsection{Classification}
Following is an analysis of the effectiveness of different supervised ML algorithms in discerning bots from humans and other types of bots. To this end, we evaluate binary classification algorithms on the binary bot dataset and multiclass classification algorithms on the multiclass MEV dataset. Note that the latter was partially generated by MEV-inspect and therefore does not generalize readily beyond the system's capabilities in a supervised setting. However, it allows us to further test our method and the utility of the features. To effectively use all labeled data, our approach deviates from the typical data science workflow. Instead of iteratively refining a model on a training subset before evaluating on a holdout set, we select methods that generally work well on small tabular datasets and evaluate them only once on the entire labeled data. This allows us to use more data to present our results but limits optimization.

For our binary and multiclass classifiers, we use the scikit-learn implementations of Random Forest and AdaBoost, and XGBoost from the package XGBoost \cite{chen_xgboost_2016}. We set the number of estimators for the Random Forest at 400 and leave the rest of the parameters at their default values. We chose these methods because they are known to work well for small tabular datasets \cite{xu_when_2021}.
Even though tree-based methods are not sensitive to scale, we apply standardization and mean-imputation in a preprocessing step.

To measure the performance of classifiers, we employ the standard binary classification metrics accuracy, recall, precision, and F1 where Bot represents the positive class. For multiclass problems, we use the macro average of the respective metric. We model the binary bot dataset with the binary version of the chosen supervised methods and the multiclass MEV dataset with their multiclass equivalent. To increase the utilization of the binary bot and multiclass MEV datasets, we use 20-fold cross validation and calculate an average of the corresponding metric, and further report a confidence interval for the mean based on the 20 values.

\begin{table}[!h]
\center
\caption{Evaluation metrics of binary classifiers on the binary bot dataset.}
\label{tab:evalset_supervised_scores}
\resizebox{\columnwidth}{!}{%
\begin{tabular}{lllll}
\toprule
{} &           Accuracy &          Precision &             Recall &                 F1 \\
Algorithm        &                    &                    &                    &                    \\
\midrule
Random Forest     &  \textbf{0.83} (0.77, 0.88) &  \textbf{0.87} (0.77, 0.97) &  0.77 (0.68, 0.86) &  0.80 (0.72, 0.88) \\
GradientBoosting &  0.82 (0.77, 0.86) &  0.85 (0.77, 0.92) &  0.78 (0.70, 0.87) &  0.80 (0.73, 0.87) \\
AdaBoost         &  \textbf{0.83} (0.78, 0.87) &  0.84 (0.75, 0.94) &  \textbf{0.80} (0.71, 0.88) &  \textbf{0.81} (0.73, 0.88) \\
\bottomrule
\end{tabular}

}
\end{table}

Testing the utility of the three classifiers Random Forest, Gradient Boosting, and AdaBoost on the binary Bot / Human problem on the binary bot dataset, we show in Table \ref{tab:evalset_supervised_scores} that the algorithms perform similarly to each other, especially considering the difference in mean accuracy and the size of the confidence intervals.

\begin{table}[!h]
\center
\caption{Evaluation metrics of multiclass classifiers on the multiclass MEV dataset.}
\label{tab:MEV_metrics_multiclass}
\resizebox{\columnwidth}{!}{%
\begin{tabular}{lllll}
\toprule
{} &           Accuracy &          Precision &             Recall &                 F1 \\
Algorithm        &                    &                    &                    &                    \\
\midrule
Random Forest     &  \textbf{0.77} (0.73, 0.81) &  \textbf{0.77} (0.72, 0.82) &  \textbf{0.77} (0.72, 0.81) &  \textbf{0.75} (0.70, 0.80) \\
GradientBoosting &  0.76 (0.72, 0.81) &  0.76 (0.71, 0.82) &  0.76 (0.70, 0.81) &  0.74 (0.68, 0.80) \\
AdaBoost         &  0.50 (0.44, 0.56) &  0.54 (0.47, 0.61) &  0.50 (0.44, 0.56) &  0.49 (0.43, 0.54) \\
\bottomrule
\end{tabular}

}
\end{table}

In Table \ref{tab:MEV_metrics_multiclass}, the performance of the surveyed methods on the four-class classification problem Arbitrage / Sandwich / Liquidation / non-MEV is displayed. Random Forest performed best and we call it the top classifier. More in-depth, for the top classifier, liquidations are the easiest to discern with an accuracy of 93\%: features capturing gas usage allow an easy distinction as liquidation addresses show unusually high values in them (as illustrated in Figure \ref{fig:feature_difference_multiclass_MEV} in the Appendix). Notably, sandwich addresses are difficult to detect with 16\% of them misclassified as Arbitrage and 13\% as non-MEV and a low accuracy of 68\%. Furthermore, 17\% of non-MEV is mistakenly classified as a sandwich bot. Again, Random Forest performs similar to Gradient Boosting, but AdaBoost performs significantly worse than in the binary Bot / Human problem. In conclusion, our features are not only suitable to distinguish general bots from humans, but also to differentiate between more specific classes of bots. While the generation of the multiclass MEV dataset does not allow for generalization beyond the capabilities of MEV-inspect, we gain the insight that our features are descriptive of various kinds of EOAs.

\subsection{Feature Investigation}
To find the most important features of the top classifier in the binary and multiclass setting and measure the impact of the most important features, we carry out further analysis based on explainable AI using the binary bot dataset and the multiclass MEV dataset.

\subsubsection{SHAP Values}

\begin{figure}[h]
  \centering
  \includegraphics[width=\columnwidth]{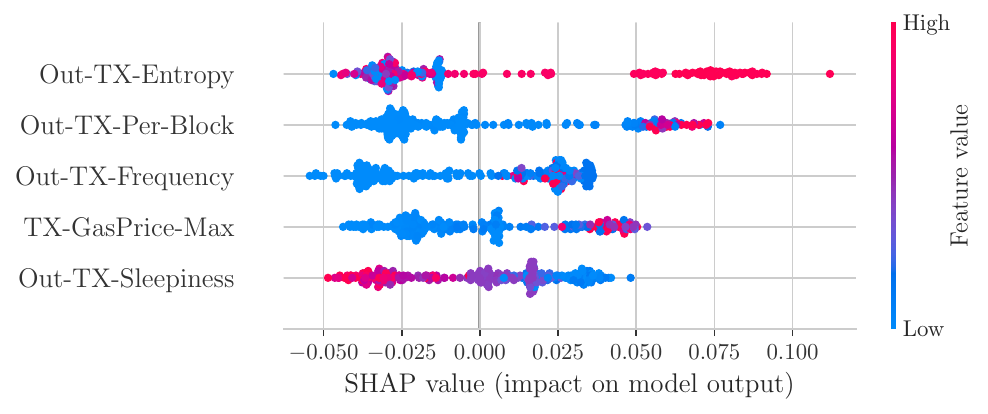}
  \caption{For each of the 5 most influential features of the top classifier a row displays 100 dots symbolizing randomly sampled addresses from the binary bot dataset. The x-axis shows each address's SHAP value, and the dot color indicates the feature value.}
  \label{fig:evalset_supervised_shap}
  \Description{For each of the 5 most influential features of the top classifier a row displays 100 dots symbolizing randomly sampled addresses from the binary bot dataset. The x-axis shows each address's SHAP value, and the dot color indicates the feature value.}
\end{figure}

SHAP (SHapley Additive exPlanations) values \cite{lundberg_unified_2017} are a method used in machine learning and data science to explain the output of a machine learning model's predictions for individual data points. They provide a way to attribute the contribution of each feature to the final prediction for a specific instance by keeping all other features the same and altering the specific feature investigated. In Figure \ref{fig:evalset_supervised_shap}, SHAP values are plotted for a sample of the binary bot dataset to further investigate the top classifier to display the most influential features. Following is a description of the features in the Figure.

\begin{enumerate}
    \item \textbf{Out-TX-Entropy}: The entropy of the distribution of times of outgoing transactions. Before calculating the entropy, the times are counted based on the hour of the day thereby yielding a discrete distribution. Higher values tend to shift the model's prediction towards the bot class.
    \item \textbf{Out-TX-Per-Block}: The average number of outgoing transactions per block. This feature signals high activity of an address and as high SHAP value addresses have high values of this feature, it seems to raise the suspicion of the model.
    \item \textbf{Out-TX-Frequency}: The average number of outgoing transactions per block in the timeframe the EOA was active. High SHAP value addresses contain both very high and very low feature value addresses. Low SHAP value addresses always have a low frequency.
    \item \textbf{TX-GasPrice-Max}: The maximum of the gas price of outgoing transactions. The higher the highest gas price paid, the more the model tends to classify an address as a bot. 
    \item \textbf{Out-TX-Sleepiness}: The GapBasedSleepiness for outgoing transactions. The more extreme the value, the greater its impact on the model's prediction. A smooth transition in color in its representation indicates that it can, by itself, shift the model's propensity towards a certain class. Compared to the other features, Out-TX-Sleepiness has low feature values for high SHAP value EOAs. This means that a low number in Out-TX-Sleepiness shifts the model's prediction to ``Bot''.
\end{enumerate}

\begin{figure}[h]
  \centering
  \includegraphics[width=\columnwidth]{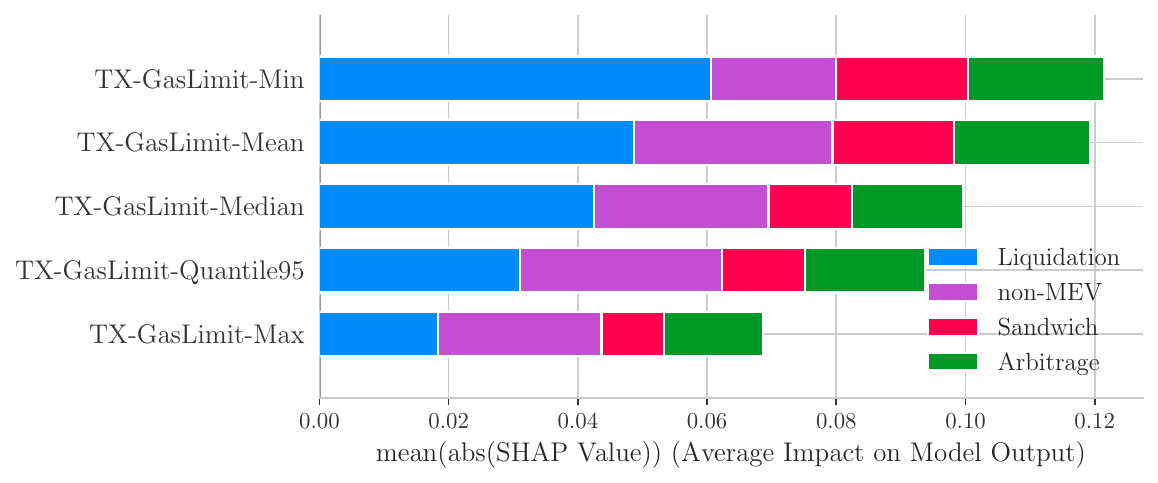}
  \caption{Stacked bar plot of the mean absolute SHAP values of all addresses in the multiclass MEV dataset with respect to the top classifier.}
  \label{fig:evalset_supervised_shap_multiclass}
  \Description{Stacked bar plot of the mean absolute SHAP values
of all addresses in the multiclass MEV dataset with respect
to the top classifier.}
\end{figure}

Figure \ref{fig:evalset_supervised_shap_multiclass} displays means of absolute SHAP values of the top classifier for all four classes in the multiclass MEV dataset. As SHAP values are defined for each class separately, they are displayed in a more aggregated view. The five most influential features are all statistics of the ``gas limit'' parameter of a transaction which defines the maximum amount of gas that may be used. Especially liquidation bots are heavily impacted by these metrics indicating they are willing to pay higher gas fees for their transactions. Gas limit metrics are least effective for identifying sandwich bots. For them, the standard deviation in Ether transferred is most significant.


\section{Discussion and Conclusions}

Our study sheds light on the role of financial bots in the Ethereum ecosystem. First, we identified and categorized the types of bots that are active on Ethereum, and we published a ground-truth dataset of 133 human and 137 bot addresses.
Second, by deploying both supervised and unsupervised machine learning models, we devised an approach to detect bots automatically. Our best-performing clustering and classification models respectively reach above 82.6\% purity and 83\% accuracy. 
Third, we used explainable AI to investigate which features are most informative to our models, finding that the most influential ones relate to time, frequency, gas price, and gas limit of outgoing
transactions. 

We note that clustering using a Gaussian Mixture Model proves to be effective in grouping certain subtypes of bots, and humans with high purity. This means that after an initial clustering step, clusters can be mapped to certain categories to single out categories of interest. Random Forest and Gradient Boosting work well both on a binary Bot/Human task and a four-class task classifying three types of MEV and non-MEV. Analyzing the features used in the model yielding the best results, we found that bots show different activity/inactivity patterns compared to humans, as captured by the \textit{GapBasedSleepiness} feature, first proposed in this work.

Further, our findings have some relevant implications.
First, our taxonomy establishes a state-of-the-art description of the existing financial bots, and therefore provides support to discern bots that can provide critical infrastructure to the DLT ecosystem from those that can be exploited for predatory trading and market manipulation.

Second, even though we acknowledge that the sample we consider is small, our study provides preliminary insights into the extent of the bot deployment in the Ethereum ecosystem.
Indeed, we found that more than 50\% of the labeled addresses are bots. 
Our work can provide additional insights into the magnitude of this phenomenon. By devising an approach to detect bots automatically, we are also able to predict with a good level of accuracy how many bots execute transactions in future incoming blocks. Similarly, in principle, our approach can be scaled to include previous transactions, allowing to provide an estimation of the magnitude of this phenomenon on the entire transaction history of Ethereum.
Third, explainable AI tools such as SHAP values can contribute to informing human users and policymakers on potential risks and characteristics of users that are likely to be bots.

Besides this, we acknowledge that our work faces some limitations and opens directions for further research. One conceptual issue is that it is hard to define exactly what a bot is, and what are the requirements to precisely define a bot based on its activity. 
The major limit to our study, however, is the small size of our labeled dataset. Our current ground-truth dataset is composed of just 270 EOAs that have been labeled by three independent annotators. A straightforward improvement to this study would be to extend the number of test blocks and annotators to augment the labeled dataset and increase the confidence level of the assigned labels. An effective strategy might be the adoption of semi-supervised learning methods \cite{zhu_semi-supervised_2005} to combine our small labeled data with readily available unlabeled data. As a restriction to a small set of blocks leads to the overrepresentation of highly active addresses, expanding the number of blocks annotated would also lead to a more representative distribution. In future work, additional data may allow for a standard data science workflow such as CRISP-DM \cite{wirth_crisp-dm_2000}, and enable optimization of model choice and parameters.
Furthermore, we currently analyze only 100,000 blocks. However, our approach is easily scalable to larger time windows and the major purpose of this paper is to provide the methodological approach to identify bots. 
Another limitation is that this study focuses on bots that engage in financial activity. However, bots are likely utilized also in other domains, for instance, to automate voting processes in Decentralized Autonomous Organizations.

Future studies could extend our taxonomy also to non-financial bots.
Further, we currently conduct multiclass classification on a restricted class of bots, i.e., MEV bots. Future work could also extend this analysis to a larger number of bot categories.
Finally, future studies could investigate the phenomenon on other blockchains, and verify whether the bot categories differ across DLTs.
\section{Acknowledgements}

This work is partially funded by the Austrian security research program KIRAS of the Federal Ministry of Finance (BMF) under the project DeFiTrace (grant agreement 905300), the FFG BRIDGE project AMALFI (grant agreement 898883), and the COMET Centre ABC (Austrian Blockchain Center) managed by the FFG (grant agreement 909237).
\vspace{2.61cm}

© {ACM} {2024}. This is the authors' version of the work. It is posted here for your personal use. Not for redistribution. The definitive Version of Record was published in {Companion Proceedings of the ACM Web Conference 2024}, https://doi.org/10.1145/3589335.3651959.
\newpage



\bibliographystyle{ACM-Reference-Format}
\bibliography{references}


\begin{thebibliography}{49}


\ifx \showCODEN    \undefined \def \showCODEN     #1{\unskip}     \fi
\ifx \showDOI      \undefined \def \showDOI       #1{#1}\fi
\ifx \showISBNx    \undefined \def \showISBNx     #1{\unskip}     \fi
\ifx \showISBNxiii \undefined \def \showISBNxiii  #1{\unskip}     \fi
\ifx \showISSN     \undefined \def \showISSN      #1{\unskip}     \fi
\ifx \showLCCN     \undefined \def \showLCCN      #1{\unskip}     \fi
\ifx \shownote     \undefined \def \shownote      #1{#1}          \fi
\ifx \showarticletitle \undefined \def \showarticletitle #1{#1}   \fi
\ifx \showURL      \undefined \def \showURL       {\relax}        \fi
\providecommand\bibfield[2]{#2}
\providecommand\bibinfo[2]{#2}
\providecommand\natexlab[1]{#1}
\providecommand\showeprint[2][]{arXiv:#2}

\bibitem[0age(2019)]%
        {0age_efficient_2019}
\bibfield{author}{\bibinfo{person}{0age}.} \bibinfo{year}{2019}\natexlab{}.
\newblock \bibinfo{title}{On {Efficient} {Ethereum} {Addresses}}.
\newblock
\newblock
\urldef\tempurl%
\url{https://medium.com/coinmonks/on-efficient-ethereum-addresses-3fef0596e263}
\showURL{%
\tempurl}


\bibitem[Auer et~al\mbox{.}(2022)]%
        {auer_miners_2022}
\bibfield{author}{\bibinfo{person}{Raphael Auer}, \bibinfo{person}{Jon Frost}, {and} \bibinfo{person}{Jose María~Vidal Pastor}.} \bibinfo{year}{2022}\natexlab{}.
\newblock \showarticletitle{Miners as intermediaries: extractable value and market manipulation in crypto and {DeFi}}.
\newblock  (\bibinfo{year}{2022}).
\newblock


\bibitem[Auer et~al\mbox{.}(2023)]%
        {auer_technology_2023}
\bibfield{author}{\bibinfo{person}{Raphael Auer}, \bibinfo{person}{Bernhard Haslhofer}, \bibinfo{person}{Stefan Kitzler}, \bibinfo{person}{Pietro Saggese}, {and} \bibinfo{person}{Friedhelm Victor}.} \bibinfo{year}{2023}\natexlab{}.
\newblock \showarticletitle{The technology of decentralized finance ({DeFi})}.
\newblock \bibinfo{journal}{\emph{Digital Finance}} (\bibinfo{date}{Aug.} \bibinfo{year}{2023}).
\newblock
\showISSN{2524-6984, 2524-6186}
\urldef\tempurl%
\url{https://doi.org/10.1007/s42521-023-00088-8}
\showDOI{\tempurl}


\bibitem[Baek et~al\mbox{.}(2019)]%
        {baek_model_2019}
\bibfield{author}{\bibinfo{person}{Hyochang Baek}, \bibinfo{person}{Junhyoung Oh}, \bibinfo{person}{Chang~Yeon Kim}, {and} \bibinfo{person}{Kyungho Lee}.} \bibinfo{year}{2019}\natexlab{}.
\newblock \showarticletitle{A {Model} for {Detecting} {Cryptocurrency} {Transactions} with {Discernible} {Purpose}}. In \bibinfo{booktitle}{\emph{2019 {Eleventh} {International} {Conference} on {Ubiquitous} and {Future} {Networks} ({ICUFN})}}. \bibinfo{pages}{713--717}.
\newblock
\urldef\tempurl%
\url{https://doi.org/10.1109/ICUFN.2019.8806126}
\showDOI{\tempurl}
\newblock
\shownote{ISSN: 2165-8536}.


\bibitem[Barczentewicz(2023)]%
        {barczentewicz_mev_2023}
\bibfield{author}{\bibinfo{person}{Mikolaj Barczentewicz}.} \bibinfo{year}{2023}\natexlab{}.
\newblock \bibinfo{title}{{MEV} on {Ethereum}: {A} {Policy} {Analysis}}.
\newblock
\newblock
\urldef\tempurl%
\url{https://doi.org/10.2139/ssrn.4332703}
\showDOI{\tempurl}


\bibitem[Benford(1938)]%
        {benford_law_1938}
\bibfield{author}{\bibinfo{person}{Frank Benford}.} \bibinfo{year}{1938}\natexlab{}.
\newblock \showarticletitle{The {Law} of {Anomalous} {Numbers}}.
\newblock \bibinfo{journal}{\emph{Proceedings of the American Philosophical Society}} \bibinfo{volume}{78}, \bibinfo{number}{4} (\bibinfo{year}{1938}), \bibinfo{pages}{551--572}.
\newblock
\showISSN{0003-049X}
\urldef\tempurl%
\url{https://www.jstor.org/stable/984802}
\showURL{%
\tempurl}
\newblock
\shownote{Publisher: American Philosophical Society}.


\bibitem[Bonifazi et~al\mbox{.}(2022)]%
        {bonifazi_defining_2022}
\bibfield{author}{\bibinfo{person}{Gianluca Bonifazi}, \bibinfo{person}{Enrico Corradini}, \bibinfo{person}{Domenico Ursino}, {and} \bibinfo{person}{Luca Virgili}.} \bibinfo{year}{2022}\natexlab{}.
\newblock \showarticletitle{Defining user spectra to classify {Ethereum} users based on their behavior}.
\newblock \bibinfo{journal}{\emph{Journal of Big Data}} \bibinfo{volume}{9}, \bibinfo{number}{1} (\bibinfo{date}{April} \bibinfo{year}{2022}), \bibinfo{pages}{37}.
\newblock
\showISSN{2196-1115}
\urldef\tempurl%
\url{https://doi.org/10.1186/s40537-022-00586-3}
\showDOI{\tempurl}


\bibitem[Cartea et~al\mbox{.}(2023)]%
        {cartea_execution_2023}
\bibfield{author}{\bibinfo{person}{Álvaro Cartea}, \bibinfo{person}{Fayçal Drissi}, {and} \bibinfo{person}{Marcello Monga}.} \bibinfo{year}{2023}\natexlab{}.
\newblock \bibinfo{title}{Execution and {Statistical} {Arbitrage} with {Signals} in {Multiple} {Automated} {Market} {Makers}}.
\newblock
\newblock
\urldef\tempurl%
\url{https://doi.org/10.2139/ssrn.4388104}
\showDOI{\tempurl}


\bibitem[Cernera et~al\mbox{.}(2023)]%
        {cernera_ready_2023}
\bibfield{author}{\bibinfo{person}{Federico Cernera}, \bibinfo{person}{Massimo La~Morgia}, \bibinfo{person}{Alessandro Mei}, \bibinfo{person}{Alberto~Maria Mongardini}, {and} \bibinfo{person}{Francesco Sassi}.} \bibinfo{year}{2023}\natexlab{}.
\newblock \showarticletitle{Ready, {Aim}, {Snipe}! {Analysis} of {Sniper} {Bots} and their {Impact} on the {DeFi} {Ecosystem}}. In \bibinfo{booktitle}{\emph{Companion {Proceedings} of the {ACM} {Web} {Conference} 2023}}. \bibinfo{publisher}{ACM}, \bibinfo{address}{Austin TX USA}, \bibinfo{pages}{1093--1102}.
\newblock
\showISBNx{978-1-4503-9419-2}
\urldef\tempurl%
\url{https://doi.org/10.1145/3543873.3587612}
\showDOI{\tempurl}


\bibitem[Chen and Guestrin(2016)]%
        {chen_xgboost_2016}
\bibfield{author}{\bibinfo{person}{Tianqi Chen} {and} \bibinfo{person}{Carlos Guestrin}.} \bibinfo{year}{2016}\natexlab{}.
\newblock \showarticletitle{{XGBoost}: {A} {Scalable} {Tree} {Boosting} {System}}. In \bibinfo{booktitle}{\emph{Proceedings of the 22nd {ACM} {SIGKDD} {International} {Conference} on {Knowledge} {Discovery} and {Data} {Mining}}} \emph{(\bibinfo{series}{{KDD} '16})}. \bibinfo{publisher}{Association for Computing Machinery}, \bibinfo{address}{New York, NY, USA}, \bibinfo{pages}{785--794}.
\newblock
\showISBNx{978-1-4503-4232-2}
\urldef\tempurl%
\url{https://doi.org/10.1145/2939672.2939785}
\showDOI{\tempurl}


\bibitem[Chen et~al\mbox{.}(2019)]%
        {chen_exploiting_2019}
\bibfield{author}{\bibinfo{person}{Weili Chen}, \bibinfo{person}{Zibin Zheng}, \bibinfo{person}{Edith C.-H. Ngai}, \bibinfo{person}{Peilin Zheng}, {and} \bibinfo{person}{Yuren Zhou}.} \bibinfo{year}{2019}\natexlab{}.
\newblock \showarticletitle{Exploiting {Blockchain} {Data} to {Detect} {Smart} {Ponzi} {Schemes} on {Ethereum}}.
\newblock \bibinfo{journal}{\emph{IEEE Access}}  \bibinfo{volume}{7} (\bibinfo{year}{2019}), \bibinfo{pages}{37575--37586}.
\newblock
\showISSN{2169-3536}
\urldef\tempurl%
\url{https://doi.org/10.1109/ACCESS.2019.2905769}
\showDOI{\tempurl}
\newblock
\shownote{Conference Name: IEEE Access}.


\bibitem[Cong et~al\mbox{.}(2021)]%
        {cong_crypto_2021}
\bibfield{author}{\bibinfo{person}{Lin~William Cong}, \bibinfo{person}{Xi Li}, \bibinfo{person}{Ke Tang}, {and} \bibinfo{person}{Yang Yang}.} \bibinfo{year}{2021}\natexlab{}.
\newblock \bibinfo{title}{Crypto {Wash} {Trading}}.
\newblock
\newblock
\urldef\tempurl%
\url{https://doi.org/10.2139/ssrn.3530220}
\showDOI{\tempurl}
\newblock
\shownote{Available at: https://ssrn.com/abstract=4529817}.


\bibitem[Daian et~al\mbox{.}(2020)]%
        {daian_flash_2020}
\bibfield{author}{\bibinfo{person}{Philip Daian}, \bibinfo{person}{Steven Goldfeder}, \bibinfo{person}{Tyler Kell}, \bibinfo{person}{Yunqi Li}, \bibinfo{person}{Xueyuan Zhao}, \bibinfo{person}{Iddo Bentov}, \bibinfo{person}{Lorenz Breidenbach}, {and} \bibinfo{person}{Ari Juels}.} \bibinfo{year}{2020}\natexlab{}.
\newblock \showarticletitle{Flash {Boys} 2.0: {Frontrunning} in {Decentralized} {Exchanges}, {Miner} {Extractable} {Value}, and {Consensus} {Instability}}. In \bibinfo{booktitle}{\emph{2020 {IEEE} {Symposium} on {Security} and {Privacy} ({SP})}}. \bibinfo{pages}{910--927}.
\newblock
\urldef\tempurl%
\url{https://doi.org/10.1109/SP40000.2020.00040}
\showDOI{\tempurl}
\newblock
\shownote{ISSN: 2375-1207}.


\bibitem[Dhillon et~al\mbox{.}(2001)]%
        {grossman_efficient_2001}
\bibfield{author}{\bibinfo{person}{Inderjit~S. Dhillon}, \bibinfo{person}{James Fan}, {and} \bibinfo{person}{Yuqiang Guan}.} \bibinfo{year}{2001}\natexlab{}.
\newblock \showarticletitle{Efficient {Clustering} of {Very} {Large} {Document} {Collections}}.
\newblock In \bibinfo{booktitle}{\emph{Data {Mining} for {Scientific} and {Engineering} {Applications}}}, \bibfield{editor}{\bibinfo{person}{Robert~L. Grossman}, \bibinfo{person}{Chandrika Kamath}, \bibinfo{person}{Philip Kegelmeyer}, \bibinfo{person}{Vipin Kumar}, {and} \bibinfo{person}{Raju~R. Namburu}} (Eds.). Vol.~\bibinfo{volume}{2}. \bibinfo{publisher}{Springer US}, \bibinfo{address}{Boston, MA}, \bibinfo{pages}{357--381}.
\newblock
\showISBNx{978-1-4020-0114-7 978-1-4615-1733-7}
\urldef\tempurl%
\url{https://doi.org/10.1007/978-1-4615-1733-7_20}
\showDOI{\tempurl}
\newblock
\shownote{Series Title: Massive Computing}.


\bibitem[Eskandari et~al\mbox{.}(2020)]%
        {eskandari_sok_2020}
\bibfield{author}{\bibinfo{person}{Shayan Eskandari}, \bibinfo{person}{Seyedehmahsa Moosavi}, {and} \bibinfo{person}{Jeremy Clark}.} \bibinfo{year}{2020}\natexlab{}.
\newblock \showarticletitle{{SoK}: {Transparent} {Dishonesty}: {Front}-{Running} {Attacks} on {Blockchain}}. In \bibinfo{booktitle}{\emph{Financial {Cryptography} and {Data} {Security}}} \emph{(\bibinfo{series}{Lecture {Notes} in {Computer} {Science}})}, \bibfield{editor}{\bibinfo{person}{Andrea Bracciali}, \bibinfo{person}{Jeremy Clark}, \bibinfo{person}{Federico Pintore}, \bibinfo{person}{Peter~B. Rønne}, {and} \bibinfo{person}{Massimiliano Sala}} (Eds.). \bibinfo{publisher}{Springer International Publishing}, \bibinfo{address}{Cham}, \bibinfo{pages}{170--189}.
\newblock
\showISBNx{978-3-030-43725-1}
\urldef\tempurl%
\url{https://doi.org/10.1007/978-3-030-43725-1_13}
\showDOI{\tempurl}


\bibitem[Flashbots(2023)]%
        {flashbots_flashbots_2023}
\bibfield{author}{\bibinfo{person}{Flashbots}.} \bibinfo{year}{2023}\natexlab{}.
\newblock \bibinfo{title}{Flashbots {Transparency} {Dashboard}}.
\newblock
\newblock
\urldef\tempurl%
\url{https://transparency.flashbots.net/}
\showURL{%
\tempurl}


\bibitem[Galletta and Pinelli(2023)]%
        {galletta_sharpening_2023}
\bibfield{author}{\bibinfo{person}{Letterio Galletta} {and} \bibinfo{person}{Fabio Pinelli}.} \bibinfo{year}{2023}\natexlab{}.
\newblock \bibinfo{title}{Sharpening {Ponzi} {Schemes} {Detection} on {Ethereum} with {Machine} {Learning}}.
\newblock
\newblock
\urldef\tempurl%
\url{http://arxiv.org/abs/2301.04872}
\showURL{%
\tempurl}
\newblock
\shownote{arXiv:2301.04872 [cs]}.


\bibitem[Haslhofer et~al\mbox{.}(2021)]%
        {haslhofer_graphsense_2021}
\bibfield{author}{\bibinfo{person}{Bernhard Haslhofer}, \bibinfo{person}{Rainer Stütz}, \bibinfo{person}{Matteo Romiti}, {and} \bibinfo{person}{Ross King}.} \bibinfo{year}{2021}\natexlab{}.
\newblock \bibinfo{title}{{GraphSense}: {A} {General}-{Purpose} {Cryptoasset} {Analytics} {Platform}}.
\newblock
\newblock
\urldef\tempurl%
\url{http://arxiv.org/abs/2102.13613}
\showURL{%
\tempurl}
\newblock
\shownote{arXiv:2102.13613 [cs]}.


\bibitem[Konstantopoulos(2022)]%
        {konstantopoulos_symbolic_2022}
\bibfield{author}{\bibinfo{person}{Georgios Konstantopoulos}.} \bibinfo{year}{2022}\natexlab{}.
\newblock \bibinfo{title}{Symbolic {MEV} {Extraction}}.
\newblock
\newblock
\urldef\tempurl%
\url{https://www.youtube.com/watch?v=VkSR9jz_C-0}
\showURL{%
\tempurl}


\bibitem[Lehar and Parlour(2021)]%
        {lehar_decentralized_2021}
\bibfield{author}{\bibinfo{person}{Alfred Lehar} {and} \bibinfo{person}{Christine~A. Parlour}.} \bibinfo{year}{2021}\natexlab{}.
\newblock \bibinfo{title}{Decentralized {Exchanges}}.
\newblock
\newblock
\urldef\tempurl%
\url{https://doi.org/10.2139/ssrn.3905316}
\showDOI{\tempurl}


\bibitem[Liu(2018)]%
        {liu_feature_2018}
\bibfield{editor}{\bibinfo{person}{Guozhu~Dong Liu, Huan}} (Ed.). \bibinfo{year}{2018}\natexlab{}.
\newblock \bibinfo{booktitle}{\emph{Feature {Engineering} for {Machine} {Learning} and {Data} {Analytics}}}.
\newblock \bibinfo{publisher}{CRC Press}, \bibinfo{address}{Boca Raton}.
\newblock
\showISBNx{978-1-315-18108-0}
\urldef\tempurl%
\url{https://doi.org/10.1201/9781315181080}
\showDOI{\tempurl}


\bibitem[Lundberg and Lee(2017)]%
        {lundberg_unified_2017}
\bibfield{author}{\bibinfo{person}{Scott Lundberg} {and} \bibinfo{person}{Su-In Lee}.} \bibinfo{year}{2017}\natexlab{}.
\newblock \bibinfo{title}{A {Unified} {Approach} to {Interpreting} {Model} {Predictions}}.
\newblock
\newblock
\urldef\tempurl%
\url{https://doi.org/10.48550/arXiv.1705.07874}
\showDOI{\tempurl}
\newblock
\shownote{arXiv:1705.07874 [cs, stat]}.


\bibitem[Marutho et~al\mbox{.}(2018)]%
        {marutho_determination_2018}
\bibfield{author}{\bibinfo{person}{Dhendra Marutho}, \bibinfo{person}{Sunarna Hendra~Handaka}, \bibinfo{person}{Ekaprana Wijaya}, {and} \bibinfo{person}{{Muljono}}.} \bibinfo{year}{2018}\natexlab{}.
\newblock \showarticletitle{The {Determination} of {Cluster} {Number} at k-{Mean} {Using} {Elbow} {Method} and {Purity} {Evaluation} on {Headline} {News}}. In \bibinfo{booktitle}{\emph{2018 {International} {Seminar} on {Application} for {Technology} of {Information} and {Communication}}}. \bibinfo{pages}{533--538}.
\newblock
\urldef\tempurl%
\url{https://doi.org/10.1109/ISEMANTIC.2018.8549751}
\showDOI{\tempurl}


\bibitem[McInnes et~al\mbox{.}(2020)]%
        {mcinnes_umap_2020}
\bibfield{author}{\bibinfo{person}{Leland McInnes}, \bibinfo{person}{John Healy}, {and} \bibinfo{person}{James Melville}.} \bibinfo{year}{2020}\natexlab{}.
\newblock \bibinfo{title}{{UMAP}: {Uniform} {Manifold} {Approximation} and {Projection} for {Dimension} {Reduction}}.
\newblock
\newblock
\urldef\tempurl%
\url{https://doi.org/10.48550/arXiv.1802.03426}
\showDOI{\tempurl}
\newblock
\shownote{arXiv:1802.03426 [cs, stat]}.


\bibitem[Mohan(2022)]%
        {mohan_automated_2022}
\bibfield{author}{\bibinfo{person}{Vijay Mohan}.} \bibinfo{year}{2022}\natexlab{}.
\newblock \showarticletitle{Automated market makers and decentralized exchanges: a {DeFi} primer}.
\newblock \bibinfo{journal}{\emph{Financial Innovation}} \bibinfo{volume}{8}, \bibinfo{number}{1} (\bibinfo{date}{Feb.} \bibinfo{year}{2022}), \bibinfo{pages}{20}.
\newblock
\showISSN{2199-4730}
\urldef\tempurl%
\url{https://doi.org/10.1186/s40854-021-00314-5}
\showDOI{\tempurl}


\bibitem[Pedregosa et~al\mbox{.}(2011)]%
        {pedregosa_scikit-learn_2011}
\bibfield{author}{\bibinfo{person}{Fabian Pedregosa}, \bibinfo{person}{Gaël Varoquaux}, \bibinfo{person}{Alexandre Gramfort}, \bibinfo{person}{Vincent Michel}, \bibinfo{person}{Bertrand Thirion}, \bibinfo{person}{Olivier Grisel}, \bibinfo{person}{Mathieu Blondel}, \bibinfo{person}{Peter Prettenhofer}, \bibinfo{person}{Ron Weiss}, \bibinfo{person}{Vincent Dubourg}, \bibinfo{person}{Jake Vanderplas}, \bibinfo{person}{Alexandre Passos}, \bibinfo{person}{David Cournapeau}, \bibinfo{person}{Matthieu Brucher}, \bibinfo{person}{Matthieu Perrot}, {and} \bibinfo{person}{Édouard Duchesnay}.} \bibinfo{year}{2011}\natexlab{}.
\newblock \showarticletitle{Scikit-learn: {Machine} {Learning} in {Python}}.
\newblock \bibinfo{journal}{\emph{Journal of Machine Learning Research}} \bibinfo{volume}{12}, \bibinfo{number}{85} (\bibinfo{year}{2011}), \bibinfo{pages}{2825--2830}.
\newblock
\showISSN{1533-7928}
\urldef\tempurl%
\url{http://jmlr.org/papers/v12/pedregosa11a.html}
\showURL{%
\tempurl}


\bibitem[Piet et~al\mbox{.}(2022)]%
        {piet_extracting_2022}
\bibfield{author}{\bibinfo{person}{Julien Piet}, \bibinfo{person}{Jaiden Fairoze}, {and} \bibinfo{person}{Nicholas Weaver}.} \bibinfo{year}{2022}\natexlab{}.
\newblock \bibinfo{title}{Extracting {Godl} [sic] from the {Salt} {Mines}: {Ethereum} {Miners} {Extracting} {Value}}.
\newblock
\newblock
\urldef\tempurl%
\url{https://doi.org/10.48550/arXiv.2203.15930}
\showDOI{\tempurl}
\newblock
\shownote{arXiv:2203.15930 [cs]}.


\bibitem[Qin et~al\mbox{.}(2021b)]%
        {qin_empirical_2021}
\bibfield{author}{\bibinfo{person}{Kaihua Qin}, \bibinfo{person}{Liyi Zhou}, \bibinfo{person}{Pablo Gamito}, \bibinfo{person}{Philipp Jovanovic}, {and} \bibinfo{person}{Arthur Gervais}.} \bibinfo{year}{2021}\natexlab{b}.
\newblock \showarticletitle{An {Empirical} {Study} of {DeFi} {Liquidations}: {Incentives}, {Risks}, and {Instabilities}}. In \bibinfo{booktitle}{\emph{Proceedings of the 21st {ACM} {Internet} {Measurement} {Conference}}}. \bibinfo{pages}{336--350}.
\newblock
\urldef\tempurl%
\url{https://doi.org/10.1145/3487552.3487811}
\showDOI{\tempurl}
\newblock
\shownote{arXiv:2106.06389 [cs, q-fin]}.


\bibitem[Qin et~al\mbox{.}(2021a)]%
        {qin_quantifying_2021}
\bibfield{author}{\bibinfo{person}{Kaihua Qin}, \bibinfo{person}{Liyi Zhou}, {and} \bibinfo{person}{Arthur Gervais}.} \bibinfo{year}{2021}\natexlab{a}.
\newblock \bibinfo{title}{Quantifying {Blockchain} {Extractable} {Value}: {How} dark is the forest?}
\newblock
\newblock
\urldef\tempurl%
\url{https://doi.org/10.48550/arXiv.2101.05511}
\showDOI{\tempurl}
\newblock
\shownote{arXiv:2101.05511 [cs]}.


\bibitem[Rabieinejad et~al\mbox{.}(2021)]%
        {rabieinejad_deep_2021}
\bibfield{author}{\bibinfo{person}{Elnaz Rabieinejad}, \bibinfo{person}{Abbas Yazdinejad}, {and} \bibinfo{person}{Reza~M. Parizi}.} \bibinfo{year}{2021}\natexlab{}.
\newblock \showarticletitle{A {Deep} {Learning} {Model} for {Threat} {Hunting} in {Ethereum} {Blockchain}}. In \bibinfo{booktitle}{\emph{2021 {IEEE} 20th {International} {Conference} on {Trust}, {Security} and {Privacy} in {Computing} and {Communications} ({TrustCom})}}. \bibinfo{pages}{1185--1190}.
\newblock
\urldef\tempurl%
\url{https://doi.org/10.1109/TrustCom53373.2021.00160}
\showDOI{\tempurl}
\newblock
\shownote{ISSN: 2324-9013}.


\bibitem[Robinson and Konstantopoulos(2020)]%
        {robinson_ethereum_2020}
\bibfield{author}{\bibinfo{person}{Dan Robinson} {and} \bibinfo{person}{Georgios Konstantopoulos}.} \bibinfo{year}{2020}\natexlab{}.
\newblock \bibinfo{title}{Ethereum is a {Dark} {Forest}}.
\newblock
\newblock
\urldef\tempurl%
\url{https://www.paradigm.xyz/2020/08/ethereum-is-a-dark-forest}
\showURL{%
\tempurl}


\bibitem[Rousseeuw(1987)]%
        {rousseeuw_silhouettes_1987}
\bibfield{author}{\bibinfo{person}{Peter~J. Rousseeuw}.} \bibinfo{year}{1987}\natexlab{}.
\newblock \showarticletitle{Silhouettes: {A} graphical aid to the interpretation and validation of cluster analysis}.
\newblock \bibinfo{journal}{\emph{J. Comput. Appl. Math.}}  \bibinfo{volume}{20} (\bibinfo{date}{Nov.} \bibinfo{year}{1987}), \bibinfo{pages}{53--65}.
\newblock
\showISSN{0377-0427}
\urldef\tempurl%
\url{https://doi.org/10.1016/0377-0427(87)90125-7}
\showDOI{\tempurl}


\bibitem[Saggese et~al\mbox{.}(2023)]%
        {saggese_assessing_2023}
\bibfield{author}{\bibinfo{person}{Pietro Saggese}, \bibinfo{person}{Esther Segalla}, \bibinfo{person}{Michael Sigmund}, \bibinfo{person}{Burkhard Raunig}, \bibinfo{person}{Felix Zangerl}, {and} \bibinfo{person}{Bernhard Haslhofer}.} \bibinfo{year}{2023}\natexlab{}.
\newblock \bibinfo{title}{Assessing the {Solvency} of {Virtual} {Asset} {Service} {Providers}: {Are} {Current} {Standards} {Sufficient}?}
\newblock
\newblock
\urldef\tempurl%
\url{https://doi.org/10.2139/ssrn.4586682}
\showDOI{\tempurl}


\bibitem[SAYADI et~al\mbox{.}(2019)]%
        {sayadi_anomaly_2019}
\bibfield{author}{\bibinfo{person}{Sirine SAYADI}, \bibinfo{person}{Sonia BEN~REJEB}, {and} \bibinfo{person}{Ziéd CHOUKAIR}.} \bibinfo{year}{2019}\natexlab{}.
\newblock \showarticletitle{Anomaly {Detection} {Model} {Over} {Blockchain} {Electronic} {Transactions}}. In \bibinfo{booktitle}{\emph{2019 15th {International} {Wireless} {Communications} \& {Mobile} {Computing} {Conference} ({IWCMC})}}. \bibinfo{pages}{895--900}.
\newblock
\urldef\tempurl%
\url{https://doi.org/10.1109/IWCMC.2019.8766765}
\showDOI{\tempurl}
\newblock
\shownote{ISSN: 2376-6506}.


\bibitem[Schwarz(1978)]%
        {schwarz_estimating_1978}
\bibfield{author}{\bibinfo{person}{Gideon Schwarz}.} \bibinfo{year}{1978}\natexlab{}.
\newblock \showarticletitle{Estimating the {Dimension} of a {Model}}.
\newblock \bibinfo{journal}{\emph{The Annals of Statistics}} \bibinfo{volume}{6}, \bibinfo{number}{2} (\bibinfo{year}{1978}), \bibinfo{pages}{461--464}.
\newblock
\showISSN{0090-5364}
\urldef\tempurl%
\url{https://www.jstor.org/stable/2958889}
\showURL{%
\tempurl}
\newblock
\shownote{Publisher: Institute of Mathematical Statistics}.


\bibitem[Shannon(1948)]%
        {shannon_mathematical_1948}
\bibfield{author}{\bibinfo{person}{C.~E. Shannon}.} \bibinfo{year}{1948}\natexlab{}.
\newblock \showarticletitle{A mathematical theory of communication}.
\newblock \bibinfo{journal}{\emph{The Bell System Technical Journal}} \bibinfo{volume}{27}, \bibinfo{number}{3} (\bibinfo{date}{July} \bibinfo{year}{1948}), \bibinfo{pages}{379--423}.
\newblock
\showISSN{0005-8580}
\urldef\tempurl%
\url{https://doi.org/10.1002/j.1538-7305.1948.tb01338.x}
\showDOI{\tempurl}
\newblock
\shownote{Conference Name: The Bell System Technical Journal}.


\bibitem[Smith(2023)]%
        {smith_maximal_2023}
\bibfield{author}{\bibinfo{person}{Corwin Smith}.} \bibinfo{year}{2023}\natexlab{}.
\newblock \bibinfo{title}{Maximal extractable value ({MEV})}.
\newblock
\newblock
\urldef\tempurl%
\url{https://ethereum.org}
\showURL{%
\tempurl}
\newblock
\shownote{Available at: https://ethereum.org/developers/docs/mev}.


\bibitem[Thibault et~al\mbox{.}(2022)]%
        {thibault_blockchain_2022}
\bibfield{author}{\bibinfo{person}{Louis~Tremblay Thibault}, \bibinfo{person}{Tom Sarry}, {and} \bibinfo{person}{Abdelhakim~Senhaji Hafid}.} \bibinfo{year}{2022}\natexlab{}.
\newblock \showarticletitle{Blockchain {Scaling} {Using} {Rollups}: {A} {Comprehensive} {Survey}}.
\newblock \bibinfo{journal}{\emph{IEEE Access}}  \bibinfo{volume}{10} (\bibinfo{year}{2022}), \bibinfo{pages}{93039--93054}.
\newblock
\showISSN{2169-3536}
\urldef\tempurl%
\url{https://doi.org/10.1109/ACCESS.2022.3200051}
\showDOI{\tempurl}
\newblock
\shownote{Conference Name: IEEE Access}.


\bibitem[Torres et~al\mbox{.}(2021)]%
        {torres_frontrunner_2021}
\bibfield{author}{\bibinfo{person}{C.~F. Torres}, \bibinfo{person}{R. Camino}, {and} \bibinfo{person}{R. State}.} \bibinfo{year}{2021}\natexlab{}.
\newblock \showarticletitle{Frontrunner {Jones} and the {Raiders} of the {Dark} {Forest}: {An} {Empirical} {Study} of {Frontrunning} on the {Ethereum} {Blockchain}}.
\newblock \bibinfo{journal}{\emph{ArXiv}} (\bibinfo{date}{Feb.} \bibinfo{year}{2021}).
\newblock
\urldef\tempurl%
\url{https://www.semanticscholar.org/paper/Frontrunner-Jones-and-the-Raiders-of-the-Dark-An-of-Torres-Camino/189c624e936060f5c106c7247ac5e87a75becdb8}
\showURL{%
\tempurl}


\bibitem[Vogelsteller and Buterin({[n.\,d.]})]%
        {vogelsteller_erc-20_nodate}
\bibfield{author}{\bibinfo{person}{Fabian Vogelsteller} {and} \bibinfo{person}{Vitalik Buterin}.} \bibinfo{year}{[n.\,d.]}\natexlab{}.
\newblock \bibinfo{title}{{ERC}-20: {Token} {Standard}}.
\newblock
\newblock
\urldef\tempurl%
\url{https://eips.ethereum.org/EIPS/eip-20}
\showURL{%
\tempurl}


\bibitem[Wang et~al\mbox{.}(2021)]%
        {wang_non-fungible_2021}
\bibfield{author}{\bibinfo{person}{Qin Wang}, \bibinfo{person}{Rujia Li}, \bibinfo{person}{Qi Wang}, {and} \bibinfo{person}{Shiping Chen}.} \bibinfo{year}{2021}\natexlab{}.
\newblock \bibinfo{title}{Non-{Fungible} {Token} ({NFT}): {Overview}, {Evaluation}, {Opportunities} and {Challenges}}.
\newblock
\newblock
\urldef\tempurl%
\url{https://doi.org/10.48550/arXiv.2105.07447}
\showDOI{\tempurl}
\newblock
\shownote{arXiv:2105.07447 [cs]}.


\bibitem[Wang et~al\mbox{.}(2022a)]%
        {wang_cyclic_2022}
\bibfield{author}{\bibinfo{person}{Ye Wang}, \bibinfo{person}{Yan Chen}, \bibinfo{person}{Haotian Wu}, \bibinfo{person}{Liyi Zhou}, \bibinfo{person}{Shuiguang Deng}, {and} \bibinfo{person}{Roger Wattenhofer}.} \bibinfo{year}{2022}\natexlab{a}.
\newblock \showarticletitle{Cyclic {Arbitrage} in {Decentralized} {Exchanges}}. In \bibinfo{booktitle}{\emph{Companion {Proceedings} of the {Web} {Conference} 2022}} \emph{(\bibinfo{series}{{WWW} '22})}. \bibinfo{publisher}{Association for Computing Machinery}, \bibinfo{address}{New York, NY, USA}, \bibinfo{pages}{12--19}.
\newblock
\showISBNx{978-1-4503-9130-6}
\urldef\tempurl%
\url{https://doi.org/10.1145/3487553.3524201}
\showDOI{\tempurl}


\bibitem[Wang et~al\mbox{.}(2022b)]%
        {wang_impact_2022}
\bibfield{author}{\bibinfo{person}{Ye Wang}, \bibinfo{person}{Patrick Zuest}, \bibinfo{person}{Yaxing Yao}, \bibinfo{person}{Zhicong Lu}, {and} \bibinfo{person}{Roger Wattenhofer}.} \bibinfo{year}{2022}\natexlab{b}.
\newblock \showarticletitle{Impact and {User} {Perception} of {Sandwich} {Attacks} in the {DeFi} {Ecosystem}}. In \bibinfo{booktitle}{\emph{{CHI} {Conference} on {Human} {Factors} in {Computing} {Systems}}}. \bibinfo{publisher}{ACM}, \bibinfo{address}{New Orleans LA USA}, \bibinfo{pages}{1--15}.
\newblock
\showISBNx{978-1-4503-9157-3}
\urldef\tempurl%
\url{https://doi.org/10.1145/3491102.3517585}
\showDOI{\tempurl}


\bibitem[Werner et~al\mbox{.}(2023)]%
        {werner_sok_2023}
\bibfield{author}{\bibinfo{person}{Sam Werner}, \bibinfo{person}{Daniel Perez}, \bibinfo{person}{Lewis Gudgeon}, \bibinfo{person}{Ariah Klages-Mundt}, \bibinfo{person}{Dominik Harz}, {and} \bibinfo{person}{William Knottenbelt}.} \bibinfo{year}{2023}\natexlab{}.
\newblock \showarticletitle{{SoK}: {Decentralized} {Finance} ({DeFi})}. In \bibinfo{booktitle}{\emph{Proceedings of the 4th {ACM} {Conference} on {Advances} in {Financial} {Technologies}}} \emph{(\bibinfo{series}{{AFT} '22})}. \bibinfo{publisher}{Association for Computing Machinery}, \bibinfo{address}{New York, NY, USA}, \bibinfo{pages}{30--46}.
\newblock
\showISBNx{978-1-4503-9861-9}
\urldef\tempurl%
\url{https://doi.org/10.1145/3558535.3559780}
\showDOI{\tempurl}


\bibitem[Wirth and Hipp(2000)]%
        {wirth_crisp-dm_2000}
\bibfield{author}{\bibinfo{person}{R. Wirth} {and} \bibinfo{person}{Jochen Hipp}.} \bibinfo{year}{2000}\natexlab{}.
\newblock \showarticletitle{{CRISP}-{DM}: {Towards} a standard process model for data mining}.
\newblock \bibinfo{journal}{\emph{Proceedings of the 4th International Conference on the Practical Applications of Knowledge Discovery and Data Mining}} (\bibinfo{date}{Jan.} \bibinfo{year}{2000}).
\newblock


\bibitem[Xu et~al\mbox{.}(2021)]%
        {xu_when_2021}
\bibfield{author}{\bibinfo{person}{Haoyin Xu}, \bibinfo{person}{Kaleab~A. Kinfu}, \bibinfo{person}{Will LeVine}, \bibinfo{person}{Sambit Panda}, \bibinfo{person}{Jayanta Dey}, \bibinfo{person}{Michael Ainsworth}, \bibinfo{person}{Yu-Chung Peng}, \bibinfo{person}{Madi Kusmanov}, \bibinfo{person}{Florian Engert}, \bibinfo{person}{Christopher~M. White}, \bibinfo{person}{Joshua~T. Vogelstein}, {and} \bibinfo{person}{Carey~E. Priebe}.} \bibinfo{year}{2021}\natexlab{}.
\newblock \bibinfo{title}{When are {Deep} {Networks} really better than {Decision} {Forests} at small sample sizes, and how?}
\newblock
\newblock
\urldef\tempurl%
\url{http://arxiv.org/abs/2108.13637}
\showURL{%
\tempurl}
\newblock
\shownote{arXiv:2108.13637 [cs, q-bio, stat]}.


\bibitem[Zhang and Chou(2023)]%
        {zhang_chi-researchsymbolic-searcher_2023}
\bibfield{author}{\bibinfo{person}{Bill Zhang} {and} \bibinfo{person}{Amy Chou}.} \bibinfo{year}{2023}\natexlab{}.
\newblock \bibinfo{title}{chi-research/symbolic-searcher}.
\newblock
\newblock
\urldef\tempurl%
\url{https://github.com/chi-research/symbolic-searcher}
\showURL{%
\tempurl}
\newblock
\shownote{original-date: 2022-09-10T16:50:58Z}.


\bibitem[Zhu(2005)]%
        {zhu_semi-supervised_2005}
\bibfield{author}{\bibinfo{person}{Xiaojin~(Jerry) Zhu}.} \bibinfo{year}{2005}\natexlab{}.
\newblock \bibinfo{booktitle}{\emph{Semi-{Supervised} {Learning} {Literature} {Survey}}}.
\newblock \bibinfo{type}{Technical {Report}}. \bibinfo{institution}{University of Wisconsin-Madison Department of Computer Sciences}.
\newblock
\urldef\tempurl%
\url{https://minds.wisconsin.edu/handle/1793/60444}
\showURL{%
\tempurl}
\newblock
\shownote{Accepted: 2012-03-15T17:19:12Z}.


\bibitem[Zwang et~al\mbox{.}(2018)]%
        {zwang_detecting_2018}
\bibfield{author}{\bibinfo{person}{Morit Zwang}, \bibinfo{person}{Shahar Somin}, \bibinfo{person}{Alex~'Sandy' Pentland}, {and} \bibinfo{person}{Yaniv Altshuler}.} \bibinfo{year}{2018}\natexlab{}.
\newblock \bibinfo{title}{Detecting {Bot} {Activity} in the {Ethereum} {Blockchain} {Network}}.
\newblock
\newblock
\urldef\tempurl%
\url{https://doi.org/10.48550/arXiv.1810.01591}
\showDOI{\tempurl}
\newblock
\shownote{arXiv:1810.01591 [cs]}.


\end{thebibliography}

\clearpage
\appendix

\section{Metrics}
\label{sec:metrics}

For a multiset of categorical variables  \( S \), we define the entropy as
\[ \text{Entropy}(S) = - \sum_{i} p_i \log_e p_i, \]
where \( p_i \) is the proportion of occurrences of the \( i \)-th category among all variables in set \( S \) 
\cite{shannon_mathematical_1948}.
For a given cluster \( C \), entropy is defined as the entropy of the multiset \( S =  \) comprising the categorical variables within \( C \).

The purity of a cluster is defined as:
\[ \text{Purity}(C) = \frac{1}{|C|} \max_i (n_i), \]
where \( |C| \) is the size of cluster \( C \) and \( n_i \) is the number of data points in \( C \) that belong to class \( i \) \cite{grossman_efficient_2001}.

\section{Clustering}
\label{sec:app_clust}

Table \ref{tab:clustering_results_purity_elbow}  shows results where the number of clusters was found automatically by minimizing the Bayesian Information Criterion (BIC) \cite{schwarz_estimating_1978} for the GMM and the elbow method for k-means because the BIC would decrease even to a high number of clusters of 100.

\begin{table}[!h]
\renewcommand\theadfont{}
\center
\caption{Representative examples and references for subcategories of bots.}
\label{tab:bot_evidence}
\resizebox{\columnwidth}{!}{

\begin{tabular}{rl}
\toprule
        \textbf{Subcategory} & \textbf{References and anecdotal evidence} \\
\midrule

       Front-running Insertion & Eskandari et. al. \cite{eskandari_sok_2020} \\ 
       Sandwich & Qin et.al. \cite{qin_quantifying_2021} \\
       Front-running Displacement & Eskandari et. al.  \cite{eskandari_sok_2020} \\ 
       Front-running Suppression & Eskandari et. al.  \cite{eskandari_sok_2020} \\ 
       Generalized front-running  & Robinson and Konstantopoulos \cite{robinson_ethereum_2020} \\
       Atomic Arbitrage & Qin et.al. \cite{qin_quantifying_2021} \\ 
       Statistical Arbitrage & Cartea et.al. \cite{cartea_execution_2023} \\

       Liquidation & Qin et.al. \cite{qin_quantifying_2021} \\

       Gen. Searching  & Konstantopoulos \cite{konstantopoulos_symbolic_2022} \\  
       
       Sniping & Cernera et. al. \cite{cernera_ready_2023} \\
       Hot Wallet & Saggese et. al. \cite{saggese_assessing_2023}
       \\  
       Deposit Wallet & Saggese et. al. \cite{saggese_assessing_2023} \\ 
       Funding \& Aggregating & Saggese et. al. \cite{saggese_assessing_2023} \\
       DEX - Custom & \url{https://github.com/bobalice7/PCS-Prediction} \\  
       Liqu. Compound & \url{https://github.com/lwYeo/Crypto-LP-Compounder} \\  

       NFT - Custom & \url{https://github.com/What-The-Commit/nft-marketplaces-offer-bot} \\  
       Parallel Minting &  \begin{tabular}[c]{@{}l@{}}
       11 consecutive transactions to minting EOAs by address \\ 0x2F3646Ef40734Ca4FE9C0201999824De14EdD823 \\ in block \url{https://etherscan.io/txs?block=14900202&p=3} \end{tabular} \\  
       Game Prog. Action & \url{https://github.com/SansegoTek/DFKQuestRunner} \\  
       Protocol Update & \url{https://etherscan.io/address/0x69f36eA1ebf4Ec9E53e3aaBF11CAF62b034ff3eE} \\  
       Rollup & \url{https://etherscan.io/address/0xA173BDdF4953C1E8be2cA0695CFc07502Ff3B1e7}
       \\ 
       Routine Paym. & \url{https://github.com/s-tikhomirov/pethreon} \\  
       Airdrop Coll. & \url{https://github.com/jaeaster/airdrop-collectooor} \\  
       Meta-wallet & \begin{tabular}[c]{@{}l@{}}
       \url{https://etherscan.io/address/0xe99c516e18241a699255Bb5317A209fa8980aE7e}  \\ \url{https://etherscan.io/address/0x5fdA118E9DbFc64DAEce3FDe657eED9333d42F1c} \\ \url{https://etherscan.io/address/0x17BF7a2f3f4758909Ba29f59824211D8286356f2} \\ \url{https://etherscan.io/address/0x3ae32939ec8d457f4528881E1E612C6534513476}
       \\
       \url{https://etherscan.io/address/0x727E98A662C20E2b47c0bDA3b1d810f1D805A200} \\  
       \end{tabular} \\  

       Non. Attr. & \url{https://etherscan.io/address/0xe36Bd1ebD771d5960fC4706d05D6cb03Ab8C3315} \\ 
       
\end{tabular}
}
\end{table}

\begin{table}[!h]
\center
\caption{Average entropy and purity of the combinations of preprocessing methods and clustering algorithms explored with a cluster size optimizing the BIC.}
\label{tab:clustering_results_purity_elbow}
\resizebox{\columnwidth}{!}{%
\begin{tabular}{lllrrr}
\toprule
       &    &      &  Purity &  Entropy &  Clusters \\
Algorithm & Imputation & \thead{Dimension. \\Reduction} &         &          &           \\
\midrule
GMM & mean & non-UMAP &   0.774 &    0.719 &        18 \\
       &    & UMAP &   0.770 &    0.726 &        29 \\
       & -1 & non-UMAP &   \textbf{0.789} &    \textbf{0.606} &        30 \\
       &    & UMAP &   0.748 &    0.765 &        29 \\
kmeans & mean & non-UMAP &   0.585 &    0.970 &         5 \\
       &    & UMAP &   0.652 &    0.923 &         5 \\
       & -1 & non-UMAP &   0.622 &    0.906 &         6 \\
       &    & UMAP &   0.704 &    0.868 &         5 \\
\bottomrule
\end{tabular}

}
\end{table}

\begin{figure}[!h]
\center
  \includegraphics[width=0.75\linewidth]{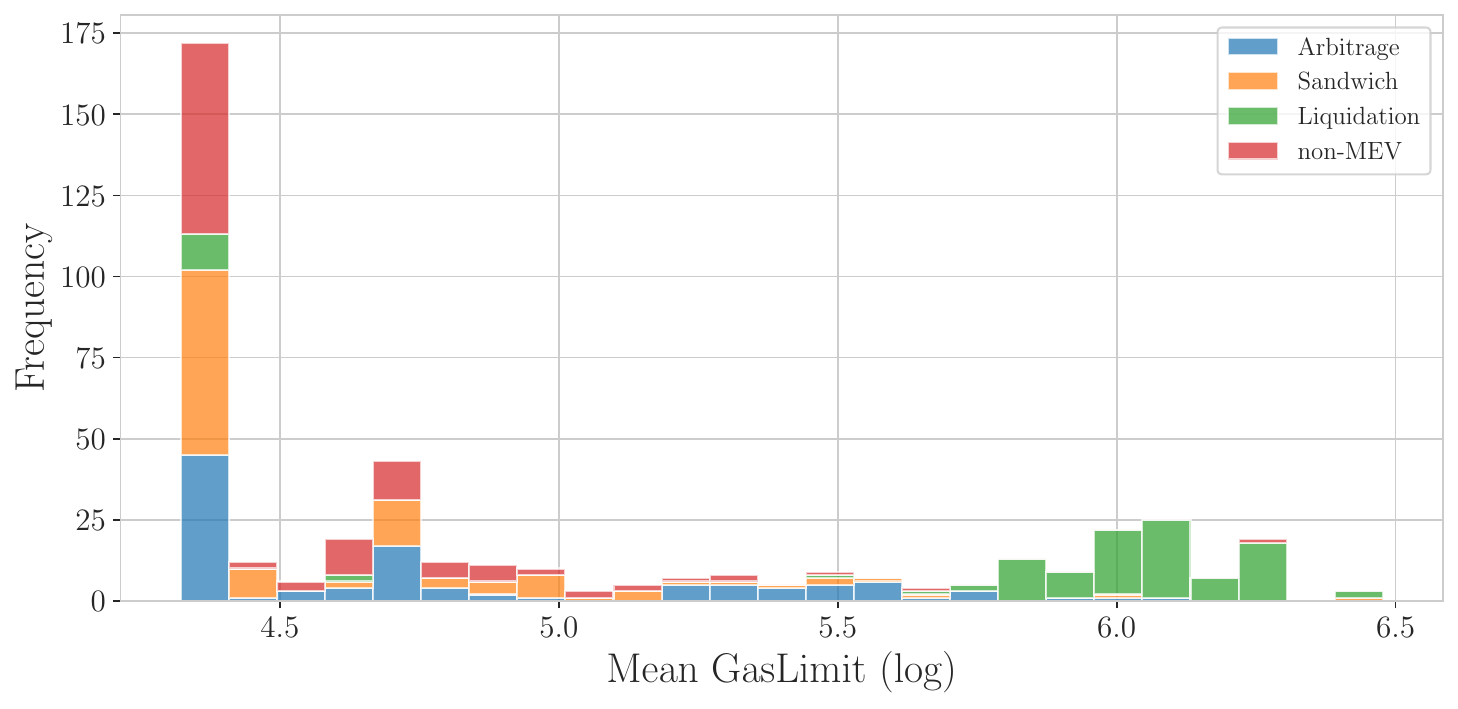}
  \caption{Log distribution of mean gasLimit of classes of the multiclass MEV dataset.}
  \label{fig:feature_difference_multiclass_MEV}
  \Description{xxx.}
\end{figure}

\begin{figure}[!h]
\center
  \includegraphics[width=\linewidth]{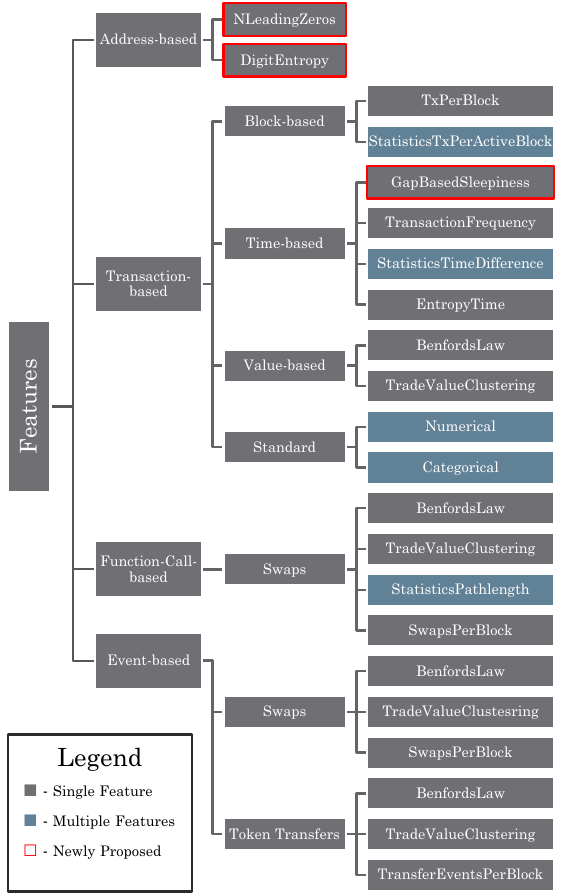}
  \caption{Overview of features.}
  \label{fig:features}
  \Description{Overview of features. The figure shows a tree where level 1 represents the categories of features and the leaves represent one or more features produced.}
\end{figure}

\newpage

\begin{table*}[!h]
\caption{Functions and methods decoded together with the data used in this work. <sender> stands for the address of the sender instead of a parameter.}
\label{tab:decoded_functions}
\resizebox{\linewidth}{!}{%
\begin{tabular}{llll}
\toprule
Name     &  Type & Signature Hash (first 8) & Relevant Parameters \\
\midrule
Transfer     &  Event & 0xddf252ad & from,to,value\\
Swap     &  Event & 0xd78ad95f  & sender,amount0In,amount1In,amount0Out,amount1Out,to \\
Swap      &  Event & 0xc42079f9 & sender,recipient,amount0,amount1\\
swapExactTokensForTokens &  Function & 0x38ed1739 & <sender>,amountIn,to,path\\
swapExactTokensForTokens &  Function & 0x472b43f3 & <sender>,amountOut,to,path\\
swapTokensForExactTokens &  Function & 0x8803dbee & <sender>,amountOut,to,path\\
swapTokensForExactETH &  Function & 0x4a25d94a & <sender>,amountOut,to,path\\
swapExactTokensForETH &  Function & 0x18cbafe5 & <sender>,amountIn,to,path\\
swapETHForExactTokens &  Function & 0xfb3bdb41 & <sender>,amountOut,to,path\\
swapExactTokensForTokensSupportingFeeOnTransferTokens &  Function & 0x5c11d795 & <sender>,amountIn,to,path\\
swapExactTokensForETHSupportingFeeOnTransferTokens &  Function & 0x791ac947 & <sender>,amountIn,to,path\\
\bottomrule
\end{tabular}

}
\end{table*}

\end{document}